\documentclass{article}

\usepackage{arxiv}

\usepackage[utf8]{inputenc}
\usepackage[T1]{fontenc}
\usepackage{hyperref}
\usepackage{url}
\usepackage{booktabs}
\usepackage{amsfonts}
\usepackage{nicefrac}
\usepackage{microtype}
\usepackage{graphicx}
\usepackage{natbib}
\usepackage{doi}

\usepackage[USenglish]{babel}
\usepackage[intlimits]{amsmath}
\usepackage{bm}
\hypersetup{colorlinks=true, citecolor=blue}

\title{Controlling the EWMA $S^2$ control chart false alarm behavior when the in-control variance level must be estimated}

\date{January 10, 2021}	

\author{ \href{https://orcid.org/0000-0002-9666-5554}{\includegraphics[scale=0.06]{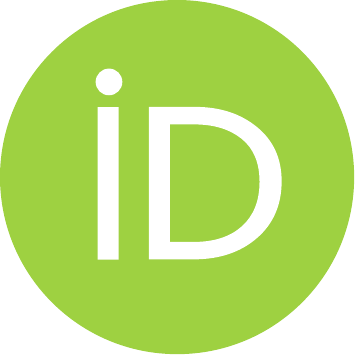}\hspace{1mm}Sven Knoth} \\
	Department of Mathematics and Statistics\\
	Helmut Schmidt University\\
	PO Box 700822\\
    22008 Hamburg, Germany\\
	\texttt{knoth@hsu-hh.de} \\
}

\renewcommand{\shorttitle}{EWMA $S^2$ for unknown $\sigma_0^2$}

\hypersetup{
	pdftitle={EWMA S2 for unknown ic level},
	pdfsubject={SPC, monitoring, surveillance, variance},
	pdfauthor={Sven Knoth},
	pdfkeywords={Control charting, S2 EWMA, phase I/II, false alarm probability}
}

\begin{document}

\maketitle

\begin{abstract}
Investigating the problem of setting control limits
in the case of parameter uncertainty is more accessible when monitoring the variance because only one parameter has to be estimated.
Simply ignoring the induced uncertainty frequently leads to control charts
with poor false alarm performances.
Adjusting the unconditional in-control (IC) average run length (ARL) makes the situation even worse.
Guaranteeing a minimum conditional IC ARL with some given
probability is another very popular approach to solving these difficulties.
However, it is very conservative as well as more complex and more difficult to communicate.
We utilize the probability of a false alarm 
within the planned number of points to be plotted on the control chart.
It turns out that adjusting this probability produces notably different
limit adjustments compared to controlling the unconditional IC ARL.
We then develop numerical algorithms to determine the respective modifications
of the upper and two-sided exponentially weighted moving average (EWMA) charts
based on the sample variance for normally distributed data.
These algorithms are made available within an \texttt{R} package.
Finally, the impacts of the EWMA smoothing constant and the size of the
preliminary sample on the control chart design and its performance are studied.
\end{abstract}	


\section{Introduction}

Applying a surveillance scheme to monitor the stability of dispersion (homogeneity, scale or other related notions)
is a common task used in industry to maintain, for example, the repeatability level of gauges,
the uniformity of certain entities over time or space, the risk level of some financial asset,
the stability of the variance underlying the control limits of a mean control chart and so forth.
To provide an explicit example, we look at a scanning electron microscope (SEM) at a semiconductor company, where
a battery of daily measurements is executed for the sake of repeatability monitoring.
Typically, well-defined features (lines, spaces and so on) on a wafer
are measured $n=5$ times, and the resulting
\begin{figure}[hbt]
\centering
\includegraphics[width=.6\textwidth]{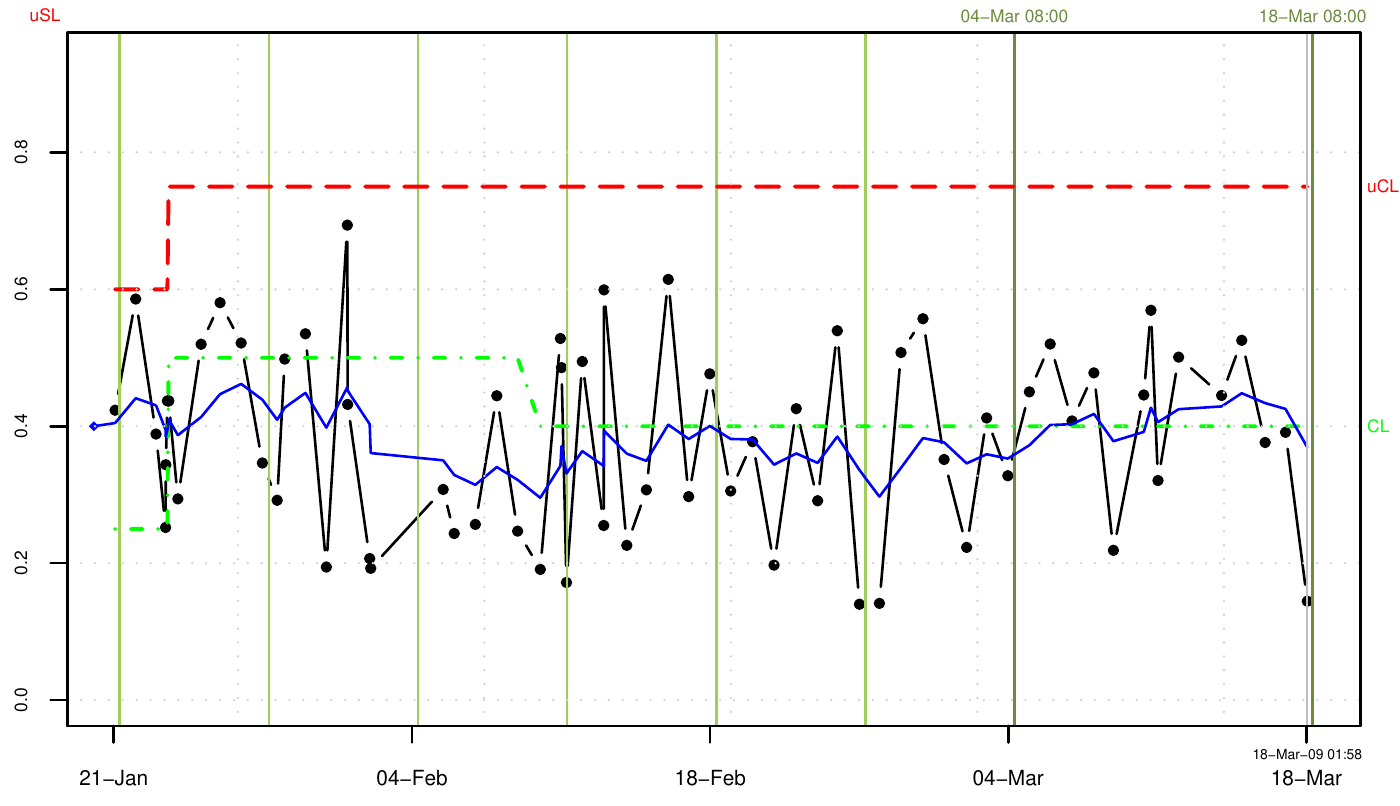}	
\caption{Shewhart $S$ chart for monitoring the short-term repeatability of a scanning electron microscope (SEM);
sample size $n=5$; ordinate scale $nm$; EWMA ($\lambda=0.2$, blue) added.}\label{fig:01}
\end{figure}
sample standard deviation is recorded on a Shewhart $S$ chart --- see Figure~\ref{fig:01}.
Hence, it is not surprising that Shewhart, cumulative sum (CUSUM) and
exponentially weighted moving average (EWMA) 
variance control charts are presented in popular textbooks, including 
\cite{Mont:2009}, pages 259, 414 and 426, respectively, and \cite{Qiu:2013}, pages 74, 146 and 198, respectively.
Here, we wish to investigate methods for calibrating EWMA schemes based on the sample variance
$S^2$ when the in-control (IC) level of the variance must be estimated based on a preliminary sample (phase I) of the IC data.
The EWMA control chart was introduced by \cite{Robe:1959} and gained much attention with and after \cite{Luca:Sacc:1990a}.
The initial works on using EWMA charts for dispersion monitoring include
\cite{Wort:Ring:1971}, \cite{Swee:1986}, \cite{Doma:Patc:1991},
\cite{Crow:Hami:1992a}, \cite{MacG:Harr:1993} and \cite{Chan:Gan:1994} --- see \cite{Knot:2005b, Knot:2010a} for more details.

With regard to EWMA charts,
\cite{Jone:Cham:Rigd:2001} started the analysis of using estimated parameters instead of merely assuming known ones.
It became, for control charts in general, an important topic in the statistical process control (SPC) literature over the course of the last 20 years, and
\cite{Jens:Jone:Cham:Wood:2006} and \cite{Psar:Vyni:Cast:2014} have provided detailed surveys on this subject.
Typically, it is assumed that the parameters are estimated through $m$ phase I samples, each of size $n$, which means that nearly all the performance measures used for control charts become uncertain.
For example, the well-known average run length (ARL; expected number of samples until the chart signals) becomes a random variable.
Two frameworks are commonly used to deal with this additional uncertainty. 
In the first framework, an unconditional form is calculated by applying a total probability mechanism. We will refer to this form as the unconditional ARL --- others use notions such as the marginal or mean ARL.
Below, we will illustrate that controlling the unconditional IC ARL induces some puzzling side effects.
The second framework started with \cite{Albe:Kall:2001, Albe:Kall:2004a, Albe:Kall:2004b}, 
who considered probabilistic bounds for performance measures, such as the conditional ARL.
More recent contributions from, for instance,
\cite{Capi:Masa:2010b},
\cite{Jone:Stei:2011}
and \cite{Gand:Kval:2013}, have stimulated
a series of additional publications employing this approach.
A popular motivation for this guarantee-a-minimum IC ARL is that it incorporates appropriately
the so-called practitioner-to-practitioner variability.
This framework has been discussed extensively for Shewhart charts monitoring the normal variance in
\cite{Eppr:Lour:Chak:2015},
\cite{Guo:Wang:2017},
\cite{Goed:EtAl:2017},
\cite{Fara:Wood:Heuc:2015},
\cite{Fara:Heuc:Sani:2018},
\cite{Apar:Mosq:Eppr:2018},
and \cite{Jardi:Sarm:Chak:Eppr:2019}.
Most of the latter works derived numerical procedures for calculating the limit adjustments, which
is much more difficult for CUSUM and EWMA control charts, where Monte Carlo simulations (bootstrapping for
the phase I dataset) are typically used. Thus, it is not surprising that nothing has been published on 
monitoring the normal variance for CUSUM and EWMA charts as of yet.
Moreover,  there are further problems with this framework. First, it is truly difficult to communicate the
probabilistic bound for the random (conditional) IC ARL to a practitioner.
Second, the modified limits are commonly quite wide, resulting in a prolongation of the detection delays.
Therefore, \cite{Capi:Masa:2020} proposed to re-estimate the modification regularly during phase II
to tighten the limits. Third, the calculations of the actual modifications for CUSUM and EWMA charts
are involved and time consuming. While the last problem will be probably be solved soon, the other problems
persist. Hence, we propose a different approach. We widen the limits of an EWMA $S^2$ chart by assuring a
certain unconditional IC run length (RL) quantile. Later, we will see that
aiming at an unconditional IC RL quantile leads to a widening of the limits, whereas deploying the
unconditional IC ARL can tighten them.

Contrary to the case of the Shewhart variance chart, there are only a few contributions dealing with EWMA variance
charts under parameter uncertainty, namely, \cite{Mara:Cast:2009}, and more recently, \cite{Zwet:Scho:Does:2015} and \cite{Zwet:Ajad:2019}.
All together control the unconditional IC ARL.
\cite{Mara:Cast:2009} investigated EWMA charts utilizing $\ln S^2$.
From their unconditional out-of-control (OOC) ARL results we pick a few, in order
to discuss what we call the unconditional ARL puzzle.
In their Table 2, unconditional OOC ARL numbers for an upper EWMA $\ln S^2$ with an
unconditional IC ARL of 370.4 were given. We provide these numbers in Table~\ref{Tab:01},
for $\lambda = 0.01$, $n = 4$ and $m \in \{10, 20, 40, 80\}$ as well as $m = \infty$ (known parameter case).
\begin{table}[hbt]
\centering
\begin{tabular}{@{}rcrrrrcr@{}} \toprule	
  $m$ & \phantom{a} & 10   &   20 &   40 &   80 & \phantom{a} & $\infty$ \\ \midrule
  ARL &             & 13.1 & 16.3 & 18.9 & 20.5 &             & 22.6 \\ \bottomrule
\end{tabular}
\caption{Side effects of using the unconditional ARL as a calibration target (IC 370.4):
Unconditional OOC ARL values (standard deviation increased by 20\%) for several
phase I sizes, $m$; sample size $n=5$; EWMA $\ln S^2$ chart with $\lambda = 0.01$.}\label{Tab:01}
\end{table}
These results include a non-remediable issue, namely, the favorable ARL values for
small $m$ suggest that small phase I samples should be utilized.
In other words, the more reliable the estimate of the unknown $\sigma_0^2$ (including in the known parameter case),
the longer one has to wait to detect this specific increase, that is, controlling the unconditional IC ARL
tightens the limits substantially, resulting in this uncommon improvement in the detection behavior. However, this
tightening greatly increases the probability of early false alarms. The heavy tail of the
unconditional IC RL distribution enlarges the corresponding mean (the unconditional IC ARL), while the
probability of low RL values becomes larger at the same time.
Later on, we will discuss this issue in more detail.
\cite{Zwet:Scho:Does:2015} utilized the unconditional ARL to adjust their limits and obtained
similar OOC ARL anomalies. Neither paper discussed these patterns.
Yet, \cite{Chak:2007} indicated that focusing on the unconditional ARL is dangerous.
In sum, controlling the unconditional IC ARL is not the way to go.

It should be added that the ARL paradigm has been criticized apart from studying the
estimation uncertainty influence on control charts limits.
For example, \cite{Yash:1985b} wrote: 
\textit{``Though ARL is	probably meaningful in the off-target situation, it can be
highly misleading when the on-target case is under study (primarily because the set of possible CUSUM paths	includes `too many' extremely `short' members)''}.
For a more recent critique, see \cite{Mei:2008} or \cite{Kuhn:Mand:Taim:2019} and the references therein.
\cite{Yash:1985b} 
also reported (for the competing CUSUM control chart):
\textit{``In general, the user of a CUSUM scheme probably feels uneasy about specifying a particular ARL for the on-target situation; what he typically wants is that the scheme will not generate a false alarm within a certain period of time (say, a shift) with probability of at least, say, 0.99.''}. This last passage refers to our design principle.

In sum, we study and propose two key features:
(i) A novel control chart design rule for incorporating estimation uncertainty that uses neither the misleading unconditional ARL nor the too conservative guarantee-a-minimum conditional ARL.
We control the unconditional false alarm probability via an unconditional IC RL quantile.
(ii)We utilize a numerical procedure that is more accurate than the Markov chain approximation \citep{Mara:Cast:2009}
and much quicker than Monte Carlo-based procedures \citep{Zwet:Scho:Does:2015}.

The paper proceeds as follows:
In Section \ref{sec:model}, we introduce the EWMA $S^2$ chart in detail and illustrate the peculiarities that emerge when
some estimate of the IC level is simply plugged in. In addition, we elucidate the deceptive
concept of adjusting the unconditional IC ARL.
Afterwards, in Section \ref{sec:algo},
we describe our novel approach and the numerical algorithm used to obtain the unconditional RL quantiles.
Eventually, in Section \ref{sec:comp}, we use this machinery to study the impact of the actual EWMA design (smoothing constant)
and of the phase I size $m$ on both the resulting control limit modification and the detection performance.
In the last section, we present our concluding discussion.

\section{EWMA $S^2$ under in-control level uncertainty} \label{sec:model}

EWMA schemes utilizing the sample variance $S^2$ are one type of EWMA chart  monitoring dispersion.
Competitors include $\ln S^2$, which is used in \cite{Crow:Hami:1992a};
$S$, as in, for example, \cite{Mitt:Stem:Tewe:1998};
the sample range $R$, which is found in \cite{Ng:Case:1989}; and
$a + b \ln ( S^2 + c )$, as in \cite{Cast:2005}.
Note that all these papers, including ours, consider normally distributed data.
There are several reasons to prefer $S^2$.
First, it is an unbiased estimator of the variance. Second, EWMA $S^2$ frequently exhibits the best detection performance -- refer to
\cite{Knot:2005b, Knot:2010a}. Third, the calculation is more feasible if all the estimation and monitoring is done with $S^2$.

Now, let $\{X_{ij}\}$ be a sequence of subgroups of independent and normally distributed data.
Each subgroup $i$ consists of $n>1$ observations $X_{i1},\ldots,X_{in}$.
As usual, we assume that the phase I data come from a stable process and that the variance change occurs at the beginning of the monitoring period or never.
Calculating the running sample variance $S_i^2$, $i=1, 2,\ldots$,
\begin{equation*}
	S_i^2 = \frac{1}{n-1} \sum_{j=1}^n \big( X_{ij} - \bar X_i \big)^2 \;\;,\quad \bar X_i = \frac{1}{n} \sum_{j=1}^n X_{ij} \,,
\end{equation*}
we feed the EWMA iteration sequence in the usual way:
\begin{equation*}
	Z_i = (1-\lambda) Z_{i-1} + \lambda S_i^2 \;\;,\quad Z_0 = z_0 = \sigma_0^2 \,.
\end{equation*}
The EWMA smoothing constant, $\lambda$, is in the
interval $(0,1]$ and controls the detection sensitivity. The EWMA sequence $\{Z_i\}$ is initialized with
the IC variance level, $\sigma_0^2$. Here, we must estimate $\sigma_0^2$ anyway.

We want to detect increases or two-sided changes in the variance level.
Hence, the following (alarm) stopping times are utilized:
\begin{align*}
  L_\text{upper} & = \min \left\{ i\ge 1: Z_i >  c_u \right\} \,, \\
  L_\text{two} & = \min \left\{ i\ge 1: Z_i >  c_u \text{ or } Z_i <  c_l \right\} \,.
\end{align*}
Note that introducing a lower reflection barrier to the upper scheme would diminish the inertia effects --- see \cite{Wood:Mahm:2005}
for more details. It would, however, also increase the complexity and dismantle the rolling estimate feature of the plain EWMA sequence $Z_i$.
Moreover, the inertia effect is less pronounced for a $S^2$-based chart with the intrinsic lower limit 0 compared to
a mean chart, which would be unbounded from below.
Therefore, we prefer the simpler design without a lower barrier.

Typically, the control limits are chosen to provide a pre-defined IC ARL, for example, by aiming for $E(L) = 500$.
In the case of a known IC variance $\sigma_0^2$, there is a rich body of literature on calculating the ARL and solving the inverse task of determining control limits for
a given IC ARL value. In this paper, we use algorithms from \cite{Knot:2005b, Knot:2007} to compute the ARL, RL quantiles,
and RL distribution  for an EWMA $S^2$ control chart. The related \texttt{R} package \texttt{spc}
offers functions that make this calculation easy.

We will start with a typical situation: Sample size $n=5$, EWMA constant $\lambda=0.1$ and target IC ARL $500$.
This setup calls for thresholds $c_l = 0.6259$ and $c_u=1.5496$ for
the two-sided EWMA alarm design, and  the threshold $c_u =1.4781$ for the upper EWMA alarm design.
For the former, we decided to use an ARL-unbiased design.
This notion was introduced by \cite{Pign:Acos:Rao:1995} and \cite{Acos:Pign:2000}, but the
phenomenon was discussed earlier in \cite{Uhlm:1982} and \cite{Krum:1992} (both in German), in \cite{Cham:Lowr:1994} and, presumably, in further publications.
For more details, refer to the more recent \cite{Knot:2010a} and \cite{Knot:Mora:2015}.
In a nutshell, ARL-unbiased designs render the ARL maximum at the IC level; in this case, $\sigma_0^2$.
For the unknown parameter case,
\cite{Guo:Wang:2017} provided results recently for ARL-unbiased Shewhart $S^2$ charts while
guaranteeing a minimum conditional IC ARL.

\begin{figure}[hbt] 
	\renewcommand{\tabcolsep}{-.7ex}
	\begin{tabular}{cc}
		\footnotesize upper & \footnotesize two-sided \\[-2ex]
		\includegraphics[width=.52\textwidth]{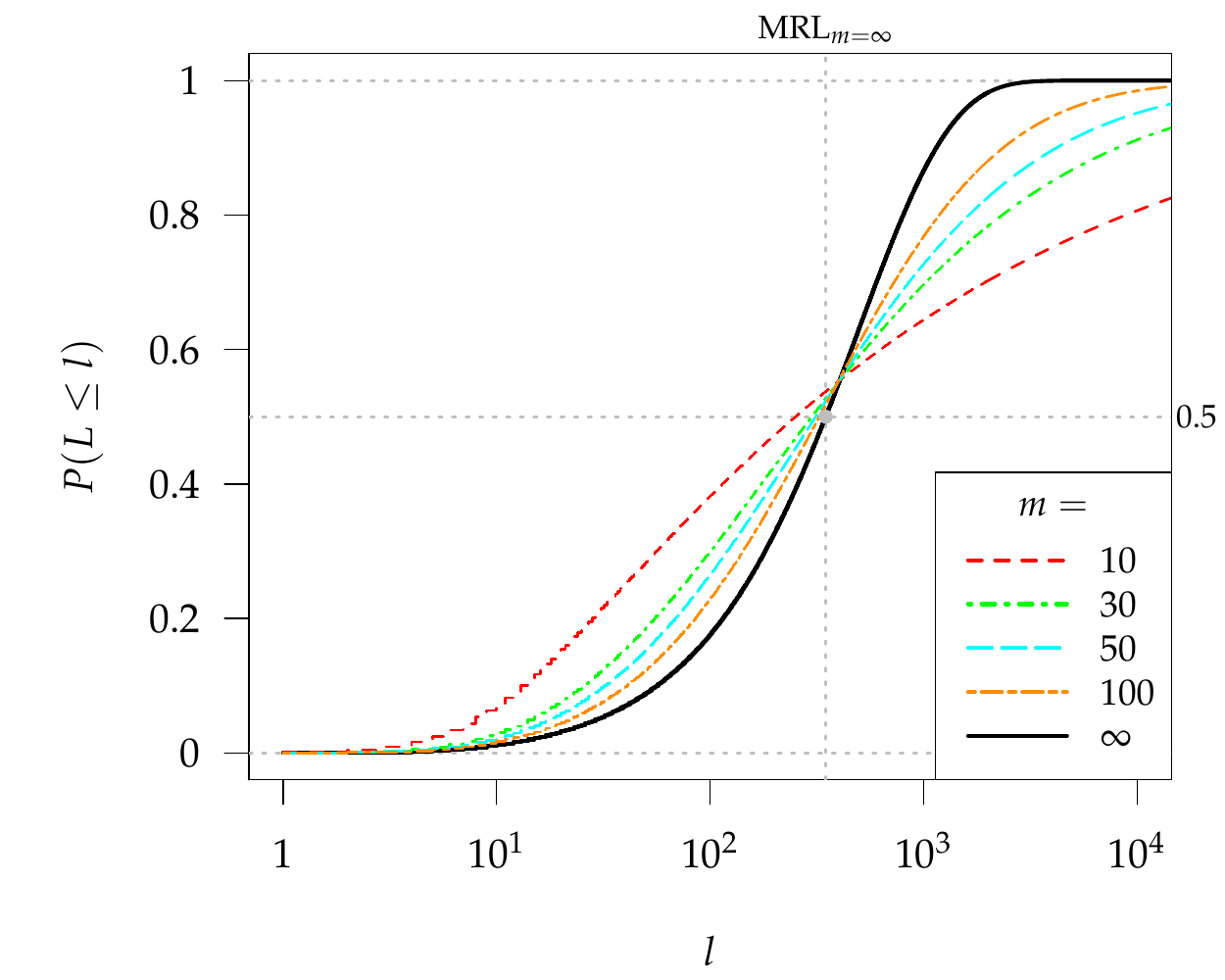} &
		\includegraphics[width=.52\textwidth]{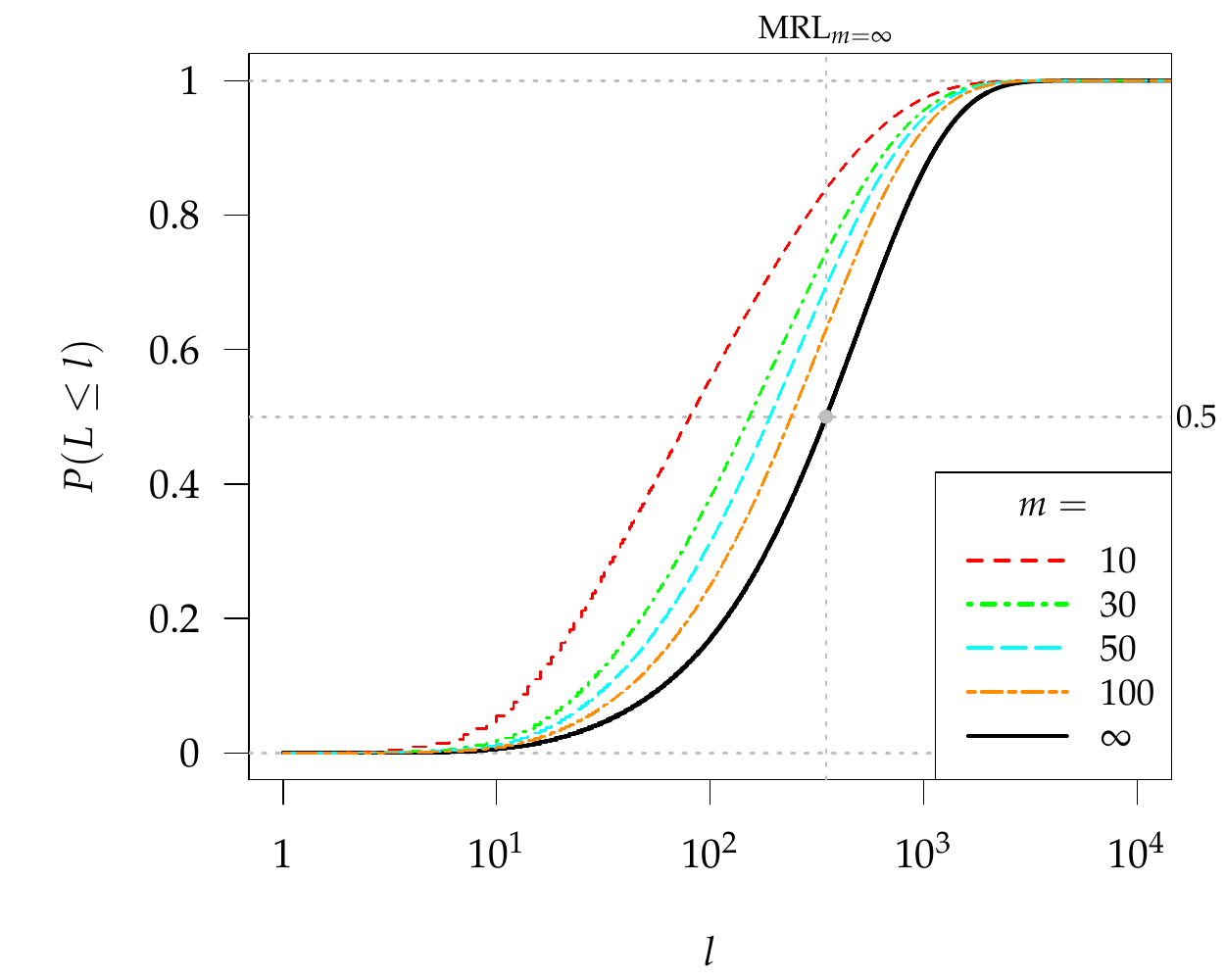}
	\end{tabular}
	\caption{Unconditional IC RL CDF for the EWMA ($\lambda=0.1$) $S^2$ ($n=5$), phase I size $m$, unadjusted case.}\label{fig:02}
\end{figure}

In this paper, we use a phase I reference dataset consisting of $m$ samples of size $n$ and
build the estimate, that is, the pooled sample variance
\begin{equation}
	\hat\sigma_0^2 = \frac{1}{m} \sum_{i=1}^m s_i^2 \label{s2pooled} \,,
\end{equation}
where $s_i^2$ denotes the sample variance of the pre-run sample $i = 1, 2, \ldots, m$.
For a discussion on appropriate estimators of the unknown $\sigma_0^2$,
we refer to \cite{Mahm:Hend:Eppr:Wood:2010}, \cite{Zwet:Scho:Does:2015} and \cite{Sale:Mahm:Jone:Zwet:Wood:2015}.
Here, we focus on the above ``natural'' estimator because it is unbiased (no further corrections are needed) and
its distribution is readily available, that is, a $\chi^2$ distribution with $m\times (n-1)$ degrees of freedom.
\cite{Zwet:Scho:Does:2015} mentioned that
\textit{“under in-control data the EWMA control charts show similar performance across all estimators,''}
where ``in-control'' refers to an uncontaminated phase I.
Replacing this “natural” estimator with a more robust \citep{Zwet:Scho:Does:2015}
or otherwise more suitable estimator does not change the framework described below, but doing so
makes the calculations more complicated. We want to emphasize, however, that all our theoretical and numerical results make
use of the pooled variance estimator \eqref{s2pooled}.
To use the aforementioned estimate means that the observed $X_{ij}$ are standardized, resulting in $\tilde{X}_{ij} = X_{ij}/\hat{\sigma}_0$.
In consequence, we run an EWMA chart design for $\sigma_0^2 = 1 = z_0$ with the limits mentioned above.
Now, we wish to study the impact on the unconditional cumulative distribution function (CDF) $P(L\le l)$ depending on the phase I sample size $m$.
Note that $P(L\le l)$ covers two sources of uncertainty: (i) phase I estimation, and (ii) phase II monitoring.
Applying the numerical algorithms described in the next section,
in Figure~\ref{fig:02}, we illustrate this CDF for several phase I sizes assuming a
known $\sigma_0^2$ during setup.
The graphs in the two-sided case feature simple patterns, namely, the smaller the phase I sample size $m$, the higher the probability that a false alarm
is flagged by $l$ for any $l \in \{1,2,\ldots\}$. In the case of the upper chart, this relation remains valid for early values
of $l \le 350$ (roughly the original RL median) only. For large values, it is
reversed. For small phase I sample sizes, such as $m \le 50$, the unconditional IC RL distribution has heavy tails.
For instance, the unconditional probability $P(L > 10^5)$ is roughly 0.1 and 0.02 for $m = 10$ and $m=30$, respectively.
Given that, for example, \cite{Zwet:Scho:Does:2015} truncate their Monte Carlo simulations at $l = 30\,000$, some potential
problems with the unconditional IC ARL and even some IC RL quantiles might be hidden.
Nonetheless, using the unconditional IC ARL to adjust the limits is dangerous because the particular tails distort the expectation and
explains the peculiar numbers in Table~\ref{Tab:01} taken from \cite{Mara:Cast:2009}. Hence, calibrating upper EWMA variance charts
by aiming for a certain unconditional IC ARL is misleading.

\section{Numerical algorithm and actual adjustment} \label{sec:algo}

We want to adjust the EWMA control limits $c_u$ and $c_l$ (just the two-sided case) in order to achieve
\begin{equation}
  P( L \le \bar{l} ) = \alpha \label{eq:design_rule} \,,
\end{equation}
where $P()$ is the unconditional IC CDF of the RL $L$.
In other words, we alter the limits so that $\bar{l}$ becomes the unconditional IC RL $\alpha$ quantile.
Recall that there is a direct link between the ARL and an RL quantile of order $\alpha$ for Shewhart charts with known IC parameters:
RL$_\alpha = \big\lceil \ln(1-\alpha)/\ln\big(1-1/\text{ARL}\big)\big\rceil$\,.
This simple formula remains approximately valid for EWMA $S^2$ charts in the IC case if $\sigma_0^2$ is known:
The median RL (MRL) is equal to $348$ and $349 \approx 347 = \lceil \ln(0.5)/\ln\big(1-1/500\big)\rceil$ with an ARL $ = 500$ for the upper and
two-sided EWMA $S^2$ charts, respectively, in Figure~\ref{fig:02}. However, this simple relationship is lost,
if we deal with the unconditional IC CDF. In the beginning of Section~\ref{sec:comp}, we provide some illustrations
of this phenomenon. This behavior is not surprising because the unconditional IC CDF is very different from the
simple geometric distribution we exploited for the Shewhart chart RL statement.
Using our rule \eqref{eq:design_rule}, we tackle the problems observed in Figure~\ref{fig:02} directly.
Appropriate choices of $\alpha$ are those that are smaller than 0.5 (we will use $\alpha  = 0.25$), while $\bar l$ could be either
derived from the ARL and RL quantile relationship for known $\sigma_0^2$ or set manually,  as in $\bar l = 1\,000$, which is used
for a typical control chart in practice. Next, we develop
an algorithm to calculate the unconditional CDF and use the result to solve the implicit
function \eqref{eq:design_rule} numerically with the secant rule.

\begin{figure}[htb]
\renewcommand{\tabcolsep}{-.7ex}
\begin{tabular}{cc}	
  \footnotesize unconditional IC SF, i.\,e. $p_{l,\text{unc.}}(1; 1, c_u) = 1 - \alpha = 0.75$& \footnotesize unconditional IC ARL \\[-2ex]
  \includegraphics[width=.52\textwidth]{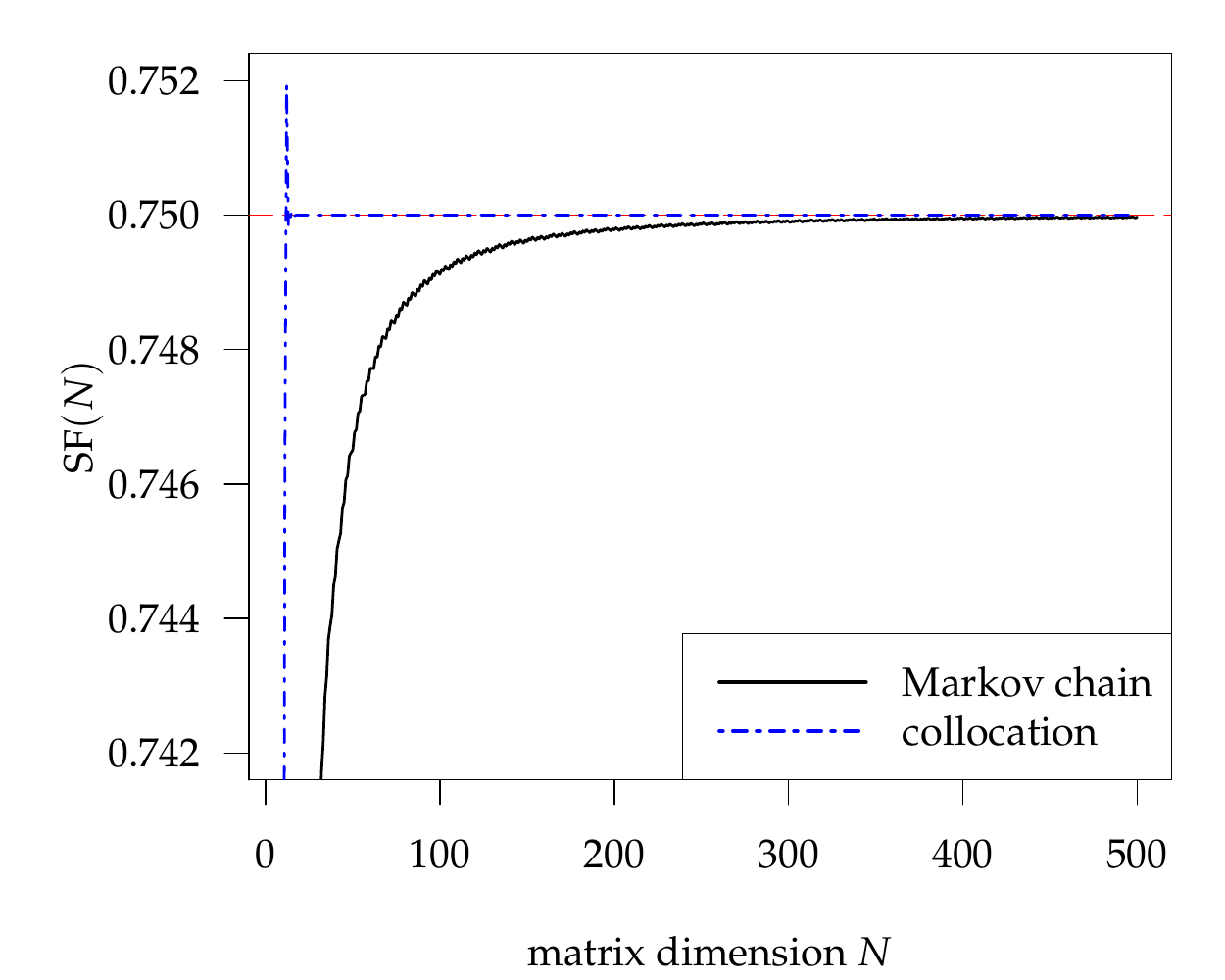} &	
  \includegraphics[width=.52\textwidth]{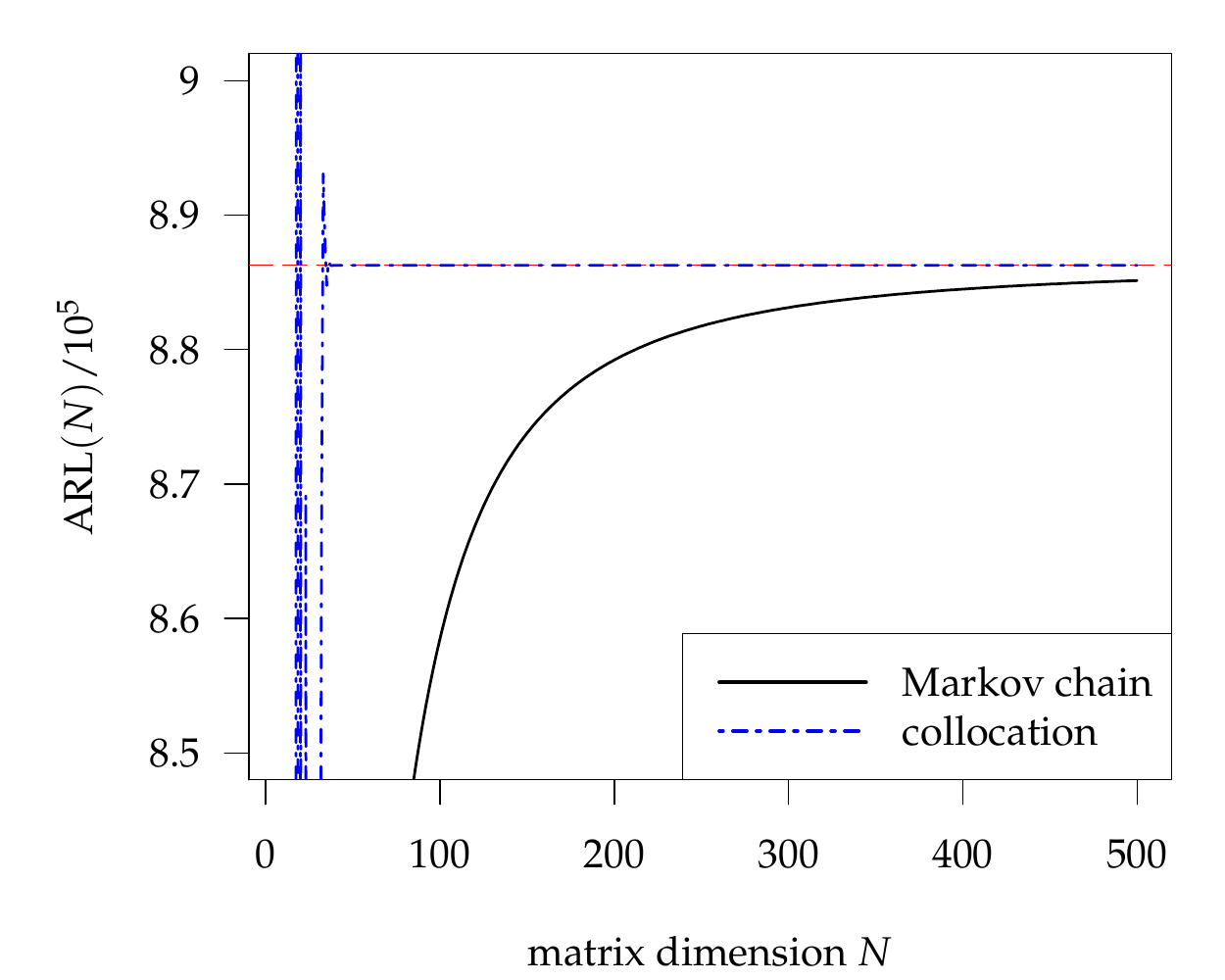}
\end{tabular}
\caption{Approximation accuracies of the Markov chain and collocation
for EWMA ($\lambda=0.1$) $S^2$ ($n=5$), phase I size $m = 50$, $P_\text{IC}(L\le 10^3) = 0.25$.}\label{fig:mchain}
\end{figure}

For calculating the unconditional CDF, we adhere to \cite{Wald:1986a}, who proposed the idea for
EWMA control charts being used to monitor a normal mean with known IC parameters.
We start with the upper chart, which only requires an adjustment to its upper limit $c_u$.
Let $p_l(z) = P(L>l\mid Z_0=z)$  denote the survival function (SF) of the RL for known $\sigma_0^2$ and starting values $Z_0 = z$.
Adding further arguments, such as the actual variance $\sigma^2$ and control limit $c_u$, we produce
its unconditional version
\begin{equation}
	p_{l,\text{unc.}}(z; \sigma^2, c_u) = \int_0^\infty f_{\hat\sigma_0^2}(s^2) p_l(z; \sigma^2 / s^2, c_u) \,ds^2
	\quad,\; l = 1, 2, \ldots \label{eq:tSF}
\end{equation}
The formula \eqref{eq:tSF} is related to (16) in \cite{Jone:Cham:Rigd:2001}. Note that only one integral is needed and that we consider the SF
instead of the probability mass function of the RL $L$.
To increase the computational speed for \eqref{eq:tSF}, the geometric tail behaviors
\citep{Wald:1986a}
of $p_l(z; \sigma^2/s^2_i, c_u)$ at each quadrature node $s^2_i$ are
(for large $l$) exploited individually. Unfortunately, it is lacking for $p_{l,\text{unc.}}(z; \sigma^2, c_u)$,
as has been mentioned previously in, for example, \cite{Psar:Vyni:Cast:2014}.
The density $f_{\hat\sigma_0^2}()$ is roughly the probability density function (PDF) of a chi-square distribution (multiply the degrees of freedom).
The numerical implementation for $p_l()$ is taken from \cite{Knot:2007}.
Because its presentation is not easily accessible, we provide some necessary details here.
We begin with the transition (from $z_0$ to $z$) density of the EWMA $S^2$ sequence as follows:
\begin{equation*}
	\delta(z_0,z) = \frac{1}{\lambda}\, f_{\chi^2;n-1}\!\left(\frac{n-1}{\sigma^2} \left[\frac{z-(1-\lambda) z_0}{\lambda}\right]\right) \frac{n-1}{\sigma^2} \,,
\end{equation*}
where $f_{\chi^2;n-1}()$ and $F_{\chi^2;n-1}()$ denote the PDF and CDF, respectively,
of a chi-square distribution with $n-1$ degrees of freedom. Then we obtain
the following recursions for the SF $p_l(z; \ldots) := p_l(z; \sigma^2 / s^2, c_u)$:
\begin{align}
	p_1(z_0;\ldots) &
	= \int_{(1-\lambda) z_0}^{c_u} \delta(z_0,z)\,dz = F_{\chi^2;n-1}\!\left(\frac{n-1}{\sigma^2} \left[\frac{c_u-(1-\lambda) z_0}{\lambda}\right]\right)
	\,, \label{eq:seq1} \\
	p_l(z_0;\ldots) &
	= \int_{(1-\lambda) z_0}^{c_u} p_{l-1}(z;\ldots)\,\delta(z_0,z)\,dz \quad,\; l=2,3,\ldots \,. \label{eq:seql}
\end{align}
A common approach to approximating the integral recursions is to replace the integrals by quadratures
\citep{Jone:Cham:Rigd:2001}. Fixed quadrature grids, however, have lower integral limit  issues since this limit, $(1-\lambda) z_0$,
depends on the argument $p_l(z_0;\ldots)$, see \cite{Knot:2005b} for a more thorough analysis.
Another idea is to apply a collocation type of procedure, as in \cite{Knot:2007}, \cite{Shu:Huan:Sub:Tsui:2013} and \cite{Huan:Shu:Jian:Tsui:2013},
who transferred the collocation principle from the ARL integral equation in \cite{Knot:2005b} to integral recursions.
Essentially, for every $l = 2, 3, \ldots$, we approximate
\begin{equation*}
	p_l(z;\ldots) \approx \sum_{s=1}^N g_{ls} \, T_s^*(z) \,, 
\end{equation*}
with $N$ suitably shifted Chebyshev polynomials $T_s^*(z)$, $s = 1, 2, \ldots, N$ (the $T_s()$ are the unit versions),
\begin{align*}
	T_s^*(z) & = T_{s-1} \big( (2z-c_u)/c_u \big) \quad,\; z \in [0,c_u] \,, \\
	T_s(z) & = \cos\big( s\,\arccos(z) \big) \quad,\;z \in[-1,1] \,.
\end{align*}
Then we pick $N$ nodes $z_r$ defined as (roots of $T_N(z)$ shifted to the interval $[0,c_u]$)
\begin{equation*}
	z_r = \frac{c_u}{2} \left[1+\cos\left( \frac{(2\,i-1)\,\pi}{2\,N} \right)\right] \quad,\; r = 1, 2, \ldots, N \,,
\end{equation*}
and consider the following recursion on the grid $\{z_r\}$, $l = 2, 3, \ldots,$
\begin{equation*}
	\sum_{s=1}^N g_{ls} \, T_s(z_r) = 
	\sum_{s=1}^N g_{l-1,s} \int_{(1-\lambda) z_r}^{c_u} T_s(z) \delta(z_r,z)\,dz \,.
\end{equation*}
These definite integrals must be determined numerically.
Because they do not depend on $l$ (only on $s$ and $r$),
we calculate them once and store them in an $N\times N$ matrix. Using this matrix and the starting vector
$\bm{g}_1 = (g_{11}, g_{12}, \ldots, g_{1N})^\prime$ derived from \eqref{eq:seq1}, we build a numerical approximation
for \eqref{eq:seql} that provides a highly accurate numerical presentation of the SF $p_l(z_0;\ldots)$ used in
\eqref{eq:tSF} to determine the unconditional CDF of the RL $L$.
The resulting SF $p_{l,\text{unc.}}(z_0; \sigma^2, c_u)$ is implemented in the \texttt{R} package \texttt{spc} as the function
\texttt{sewma.sf.prerun(l, lambda, 0, cu, sigma, n-1, m*(n-1), hs=z0, sided="upper")}, see the Appendix for an example of an application.
In Figure~\ref{fig:mchain}, we compare the approximation accuracies of the collocation and the Markov chain framework.
We investigate the EWMA $S^2$ with $\lambda=0.1$ and sample size $n = 5$ and set $\bar{l}=10^3$ and $\alpha = 0.25$, resulting in $c_u = 1.719846$.
The integral in \eqref{eq:tSF} is approximated by the Gau\ss{}-Legendre quadrature with 60 nodes,
while replacing the upper limit $\infty$ by 1.773 ($1 - 10^{-10}$ quantile of a chi-square distribution
with $m\times(n-1) = 200$ degrees of freedom divided by 200).
This integral is deployed for the SF $p_{l,\text{unc.}}(z_0; \sigma^2, c_u)$ and the unconditional ARL as well.
The matrix dimension $N$ indicates the size of the collocation basis (see above) and the number of
transient states of the Markov chain.
From the two figures, we conclude that collocation with $N = 50$ yields much higher accuracy than
the Markov chain with $N = 500$. For calculating $p_{l,\text{unc.}}(1; 1, c_u)$,
collocation needs about 1 second for $N = 50$, while the Markov chain approximation
requires 3, 6, 15 and 22 seconds for $N = 200$, $300$, $400$ and $500$, respectively.
Eventually, we want to solve \eqref{eq:design_rule} as an implicit, continuous function of the upper limit $c_u$
numerically by executing a secant rule-type algorithm. The starting values will be slightly increased limits from the
known parameter case.

Turning to the two-sided case, we face additional problems. First, the numerical procedure (collocation proceeds ``piece-wise'' now) becomes
more involved and therefore more time consuming.
The general idea follows what has been outlined above, so we will skip the details \citep[for an elaborated description see][]{Knot:2005b}.
The second problem is that we must now determine two limits, $c_l$ and $c_u$, without an intrinsic symmetric limit design, unlike what
we encountered for monitoring the normal mean. Hence, we must either deploy the symmetric design $\sigma_0^2 \pm c$ by simply ignoring
the asymmetric behavior of EWMA $S^2$ or enforce something similar to the ARL-unbiased design used in the known-parameter case.
We try two concepts: (i) make the unconditional $P_\sigma(L\le \bar{l})$
minimal in $\sigma=\sigma_0=1$ (with no loss of generality) --- denoted henceforth as the
``unbiased'' version, and (ii) perform (i) for the known parameter case (much faster) and expand the resulting limits $(c_l^\infty, c_u^\infty)$
by incorporating the correction $\xi > 1$ via $c_l = c_l^\infty / \xi$ and $c_u = c_u^\infty\cdot \xi$ so that we achieve
the unconditional $P_\text{IC}(L\le \bar{l}) = \alpha$ --- we label this method as the ``quasi-unbiased'' method. 
\begin{figure}[htb]
\renewcommand{\tabcolsep}{-.7ex}
\begin{tabular}{cc}	
  \footnotesize unconditional $P_\sigma(L\le 10^3)$ & \footnotesize unconditional ARL \\[-2ex]
  \includegraphics[width=.52\textwidth]{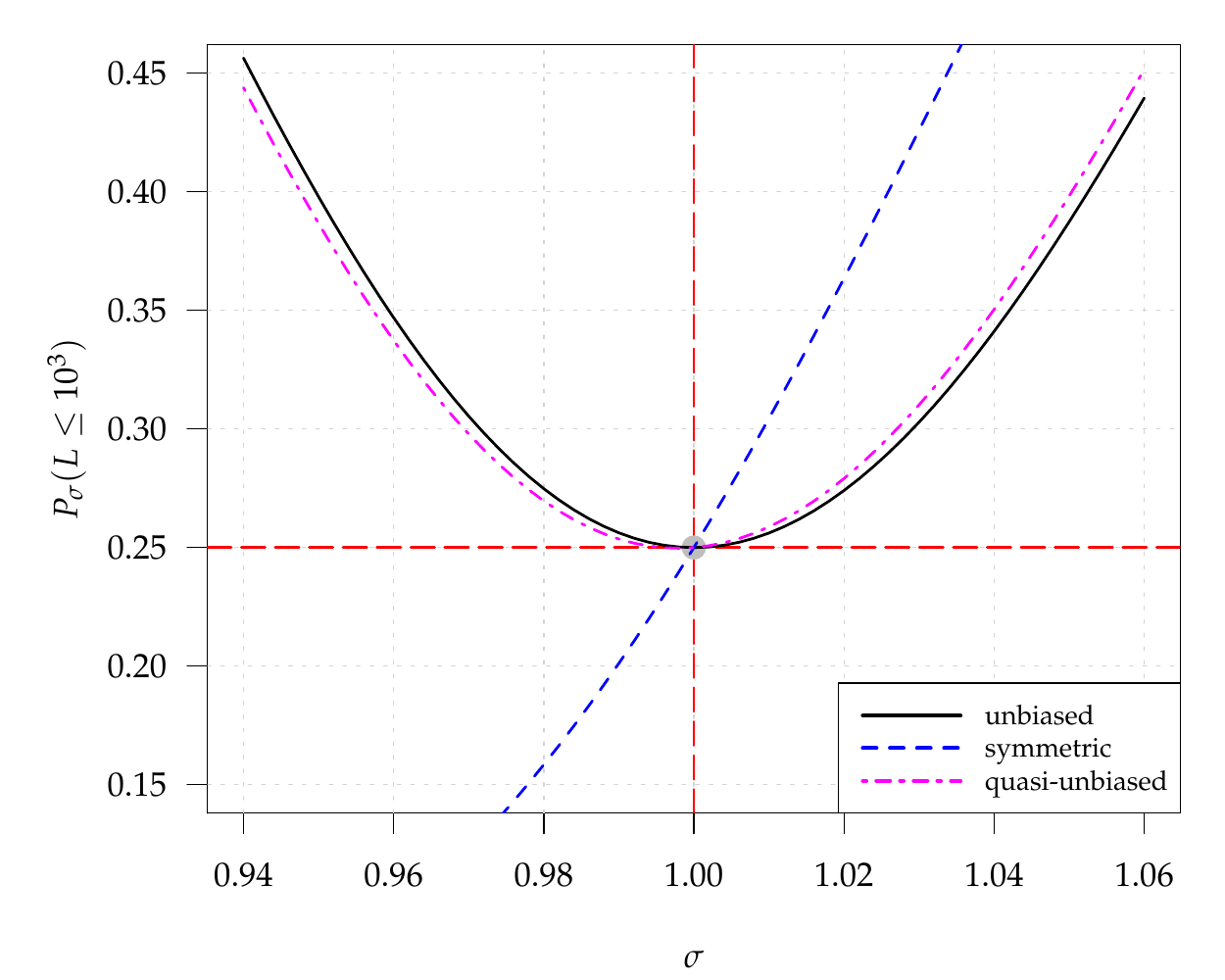} &	
  \includegraphics[width=.52\textwidth]{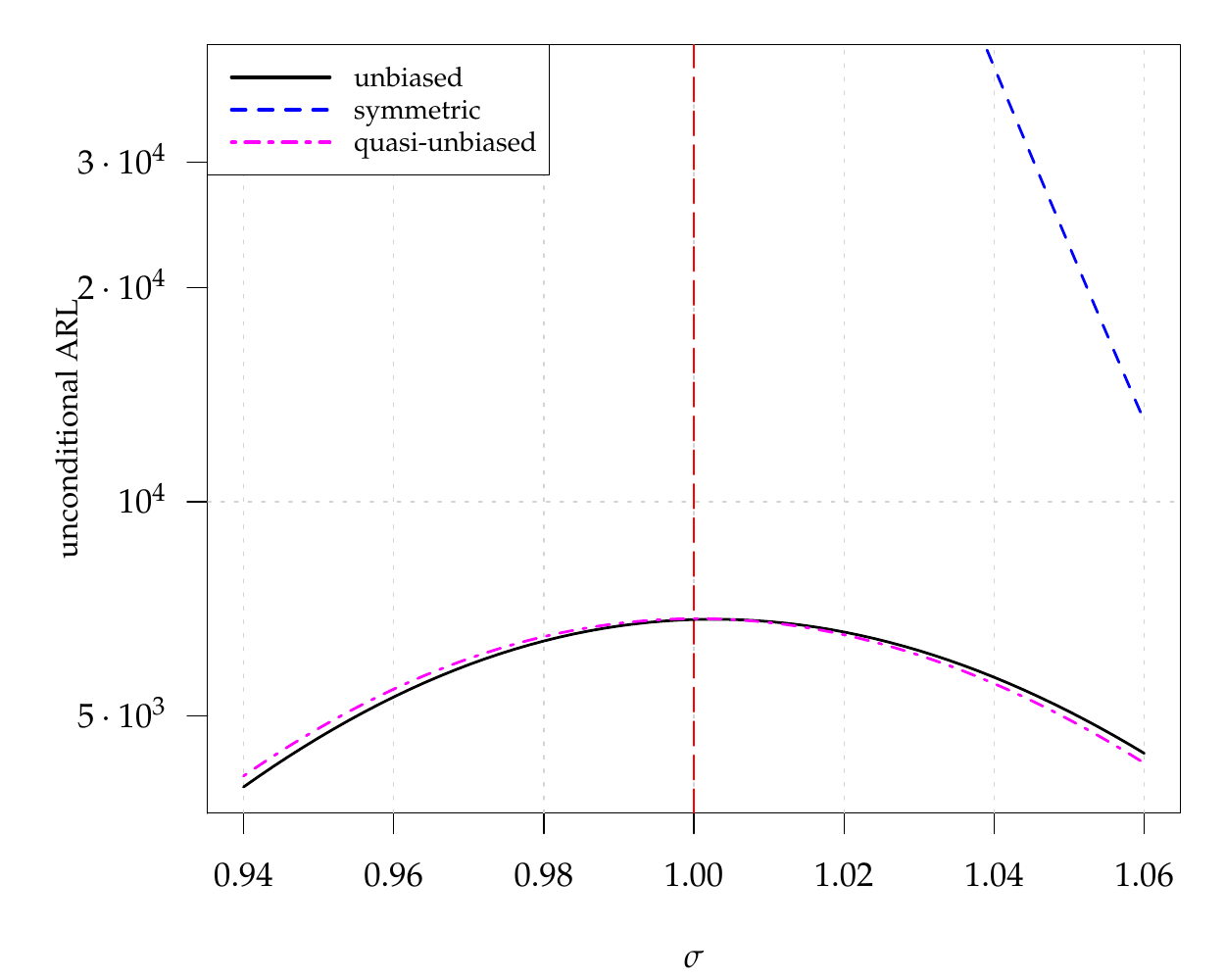}
\end{tabular}
\caption{Judging ``unbiasedness'' in the two-sided case: The unconditional $P_\sigma(L\le 10^3)$
and ARL as functions of the actual standard deviation $\sigma$
for EWMA ($\lambda=0.1$) $S^2$ ($n=5$), phase I size $m = 50$, $P_\text{IC}(L\le 10^3) = 0.25$.}\label{fig:two_sided}
\end{figure}
All three approaches
are illustrated in the following example: $\lambda = 0.1$, $\bar l = 10^3$, $\alpha = 0.25$, $n = 5$ and $m = 50$.
In all cases, we determine the new limits numerically by essentially applying the secant rule
(a more sophisticated implementation is the function \texttt{uniroot()} in \texttt{R}).
The resulting limits are
$(0.280153, 1.719847)$ (half width $c = 0.719847$),
$(0.528670, 1.824855)$ and
$(0.526394, 1.817301)$ (correction factor $\xi=1.065821$ applied to
$(c_l^\infty, c_u^\infty)=(0.561042, 1.705071)$) for the
symmetric, unbiased and quasi-unbiased approaches, respectively.
In Figure~\ref{fig:two_sided},
we illustrate the resulting profiles for $P(L\le 10^3)$ and the ARL as functions of the actual standard deviation $\sigma$.
For both, we deployed the unconditional distribution. Note that the simple symmetric design exhibits profiles (SF and ARL)
that are far from being unbiased. Moreover, the unconditional OOC ARL is very large for $\sigma < \sigma_0 = 1$ (the
lower limit is much smaller than those of the two competitors).
Hence, from this point forward, we will drop the symmetric limit design. The more sophisticated procedures feature rather equal profiles.
In the sequel, we will apply the unbiased approach to be on the safe side. However, because it needs considerably more computing time
than the quasi-unbiased scheme, we recommend the latter for daily practice.
We should note that the large unconditional ARL values are the result of the special setup utilized here. For instance,
when $\sigma_0^2$ is known, we observe an IC ARL of about
$3\,450$, which is then inflated to about 
$6\,800$ by two sources: the widened limits and enlarged tails of the unconditional RL distribution. To achieve smaller
values, $\bar l = 10^3$ should be decreased or $\alpha = 0.25$ should be increased.
It should be noted that that using $\sigma_0 = 1$ does not violate the generality of our results.
Hence, $\sigma$ will refer to the standardized version $\sigma_0 = 1$. For example, $\sigma = 1.2$ means that
the OOC standard deviation is 20\% larger than its (unknown) IC counterpart.
Finally, we should emphasize that the unconditional $\alpha=0.25$ RL quantile $\bar{l} = 10^3$
differs substantially from measures such as ``$\text{\textit{Percentile}}_\text{marginal}$'' \citep{Zhan:Mega:Wood:2014},
where the RL quantile for known $\sigma_0^2$ replaces $p_l(z;\ldots)$ in \eqref{eq:tSF}.
This weighted average over all conditional RL quantiles is much larger. For example, we
obtain 244\,325 for $\alpha=0.25$.
The exception is the unconditional ARL. To calculate it, we could utilize either
\eqref{eq:tSF}  and plug in the conditional means
or sum up $p_{l,\text{unc.}}(z;\ldots)$ over all $l$, which is just the expectation of the unconditional
RL distribution.
It remains somewhat unclear what exactly is being measured with $\text{\textit{Percentile}}_\text{marginal}$.

After deriving these quite involved algorithms, we use them to illustrate the dependence of $c_u$ on the phase I size $m$.
\begin{figure}[hbt] 
\centering	
\includegraphics[width=.55\textwidth]{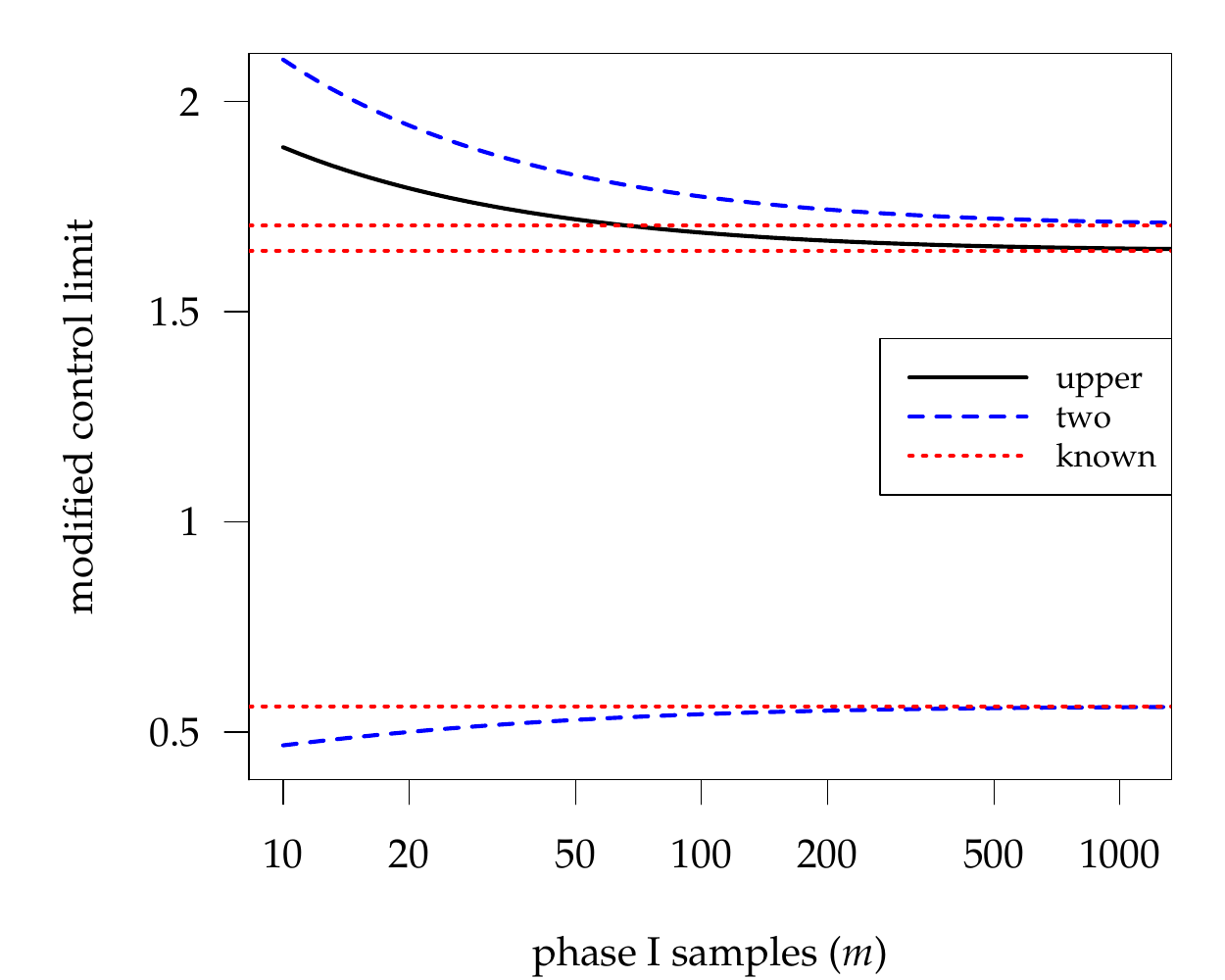}		
\caption{Modified control limits for $P_\text{IC}(L\le 10^3) = 0.25$, upper and two-sided EWMA ($\lambda=0.1$) $S^2$ ($n=5$),
$m$ varies.}\label{fig:03}
\end{figure}
Utilizing our setup with $\bar{l} = 1\,000$, $\alpha = 0.25$  and EWMA's $\lambda = 0.1$,
we start with the limits for known $\sigma_0^2$ as a benchmark ---
$c_u = 1.6453$ and $(c_l=0.5610, c_u=1.7051)$ for the upper and two-sided cases, respectively.
For realistic values of phase I sample sizes $m$ between 10 and 1\,000, we obtain widened limits, as can be seen in Figure~\ref{fig:03}.
From the profiles, we conclude that the widening is less pronounced than might be expected. From sizes $m=50$ on, the resulting limits on the
control chart device in use would not really differ from the ideal case in which $\sigma_0^2$ is known.

\begin{figure}[hbt] 
\renewcommand{\tabcolsep}{-.7ex}
\begin{tabular}{cc}	
  \footnotesize upper & \footnotesize two-sided \\[-2ex]
  \includegraphics[width=.52\textwidth]{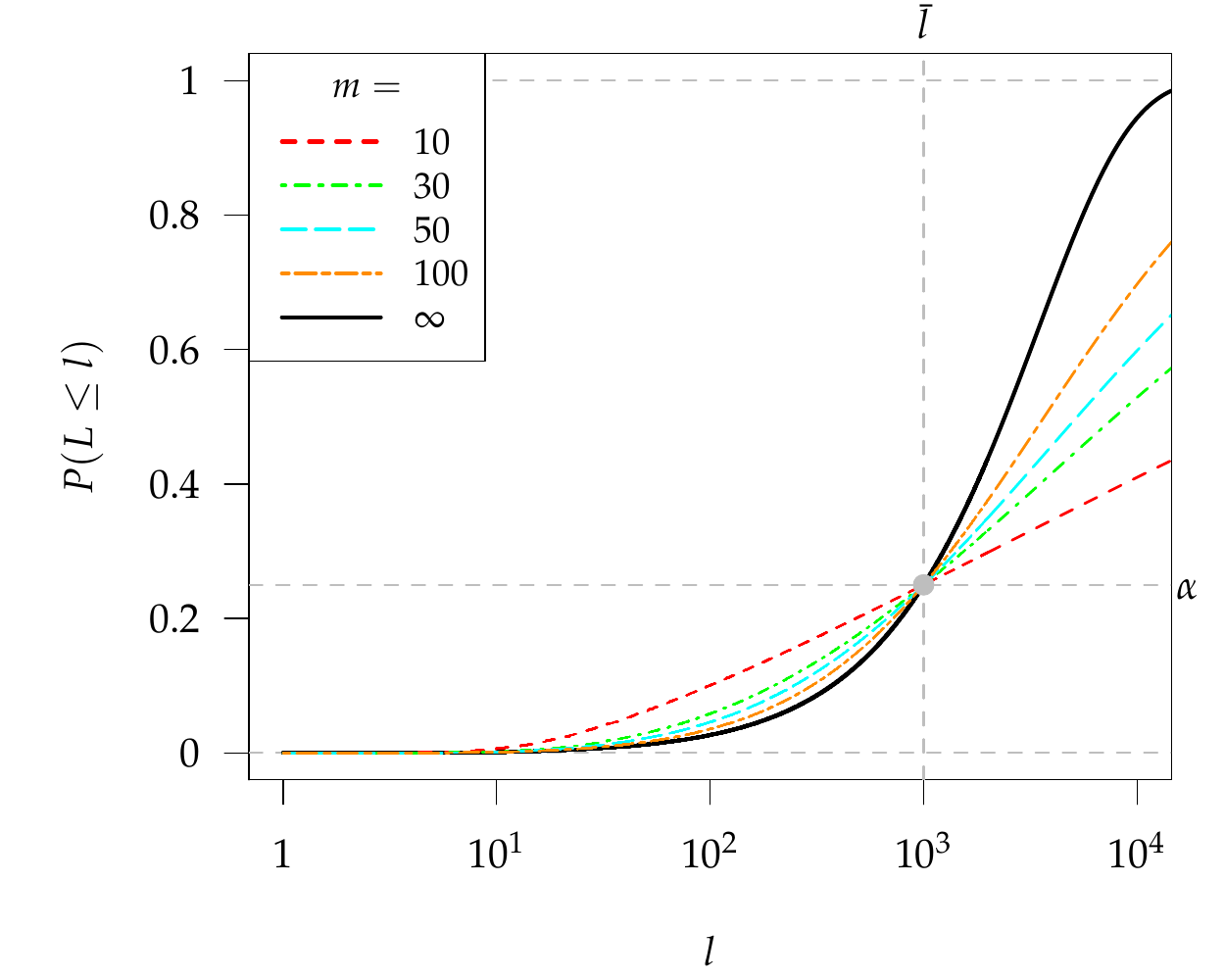} &	
  \includegraphics[width=.52\textwidth]{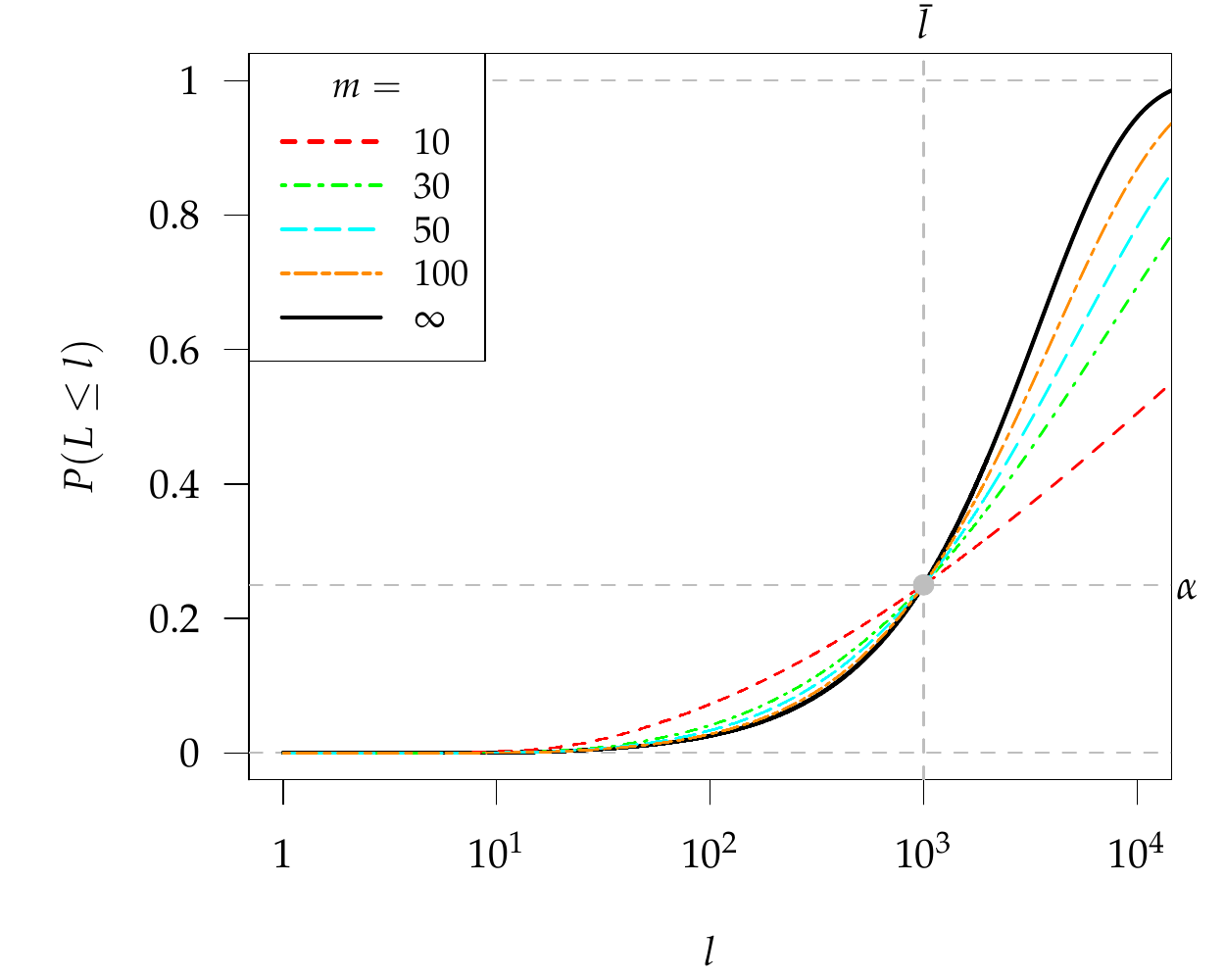}
\end{tabular}
\caption{Unconditional IC RL CDF for EWMA ($\lambda=0.1$) $S^2$ ($n=5$), phase I size $m$, $P_\text{IC}(L\le 10^3) = 0.25$.}\label{fig:04}
\end{figure}

Applying these new control limits changes the CDF profiles from those in Figure~\ref{fig:02}
to the ones presented in Figure~\ref{fig:04}.
All profiles go through the point $(\bar{l},\alpha)$ by construction, of course. However, we observe that the smaller the phase size $m$, the more likely the very early false alarms.

\begin{figure}[hbt] 
	\renewcommand{\tabcolsep}{-.7ex}
	\begin{tabular}{cc}	
		\footnotesize upper & \footnotesize two-sided \\[-2ex]
		\includegraphics[width=.52\textwidth]{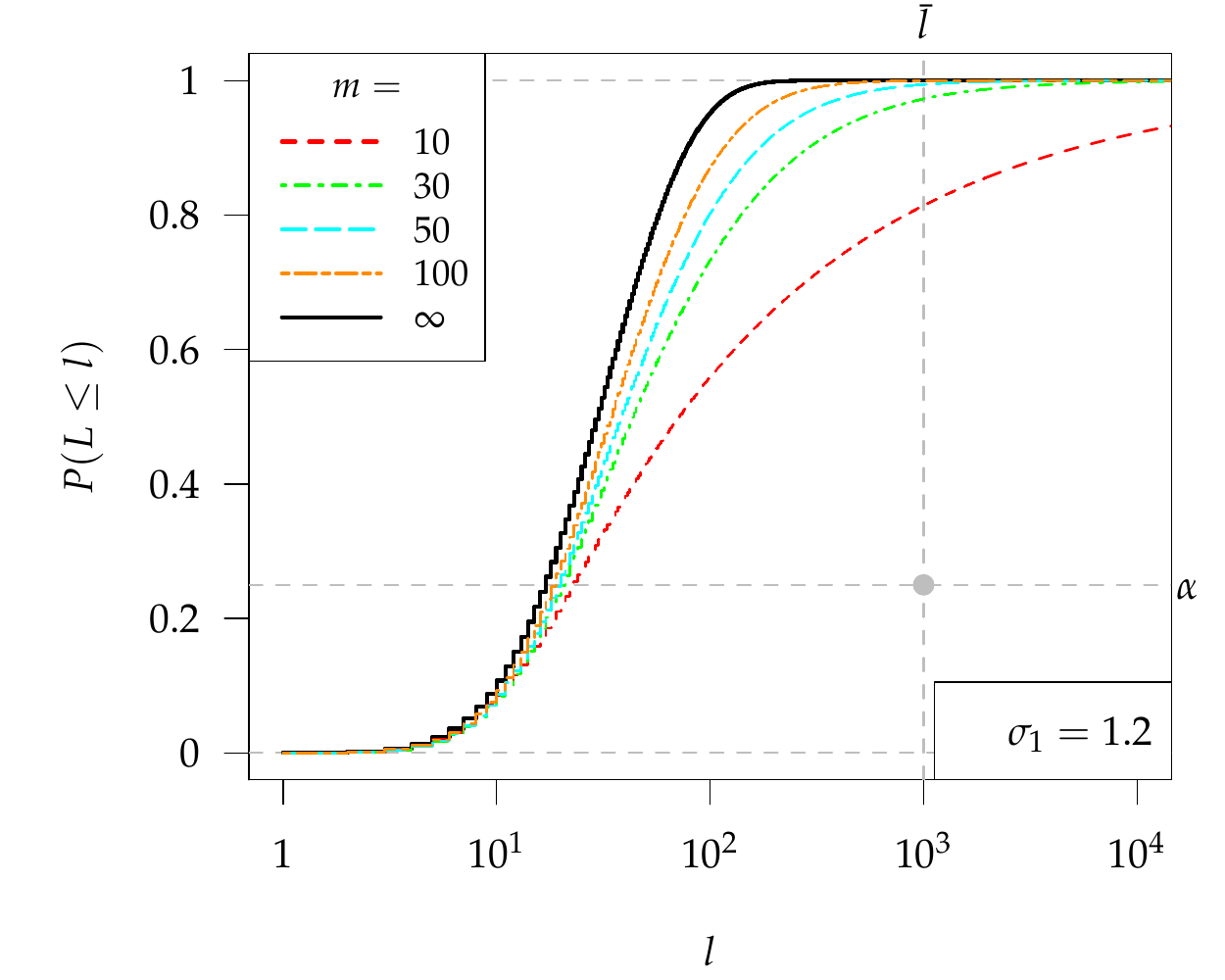} &	
		\includegraphics[width=.52\textwidth]{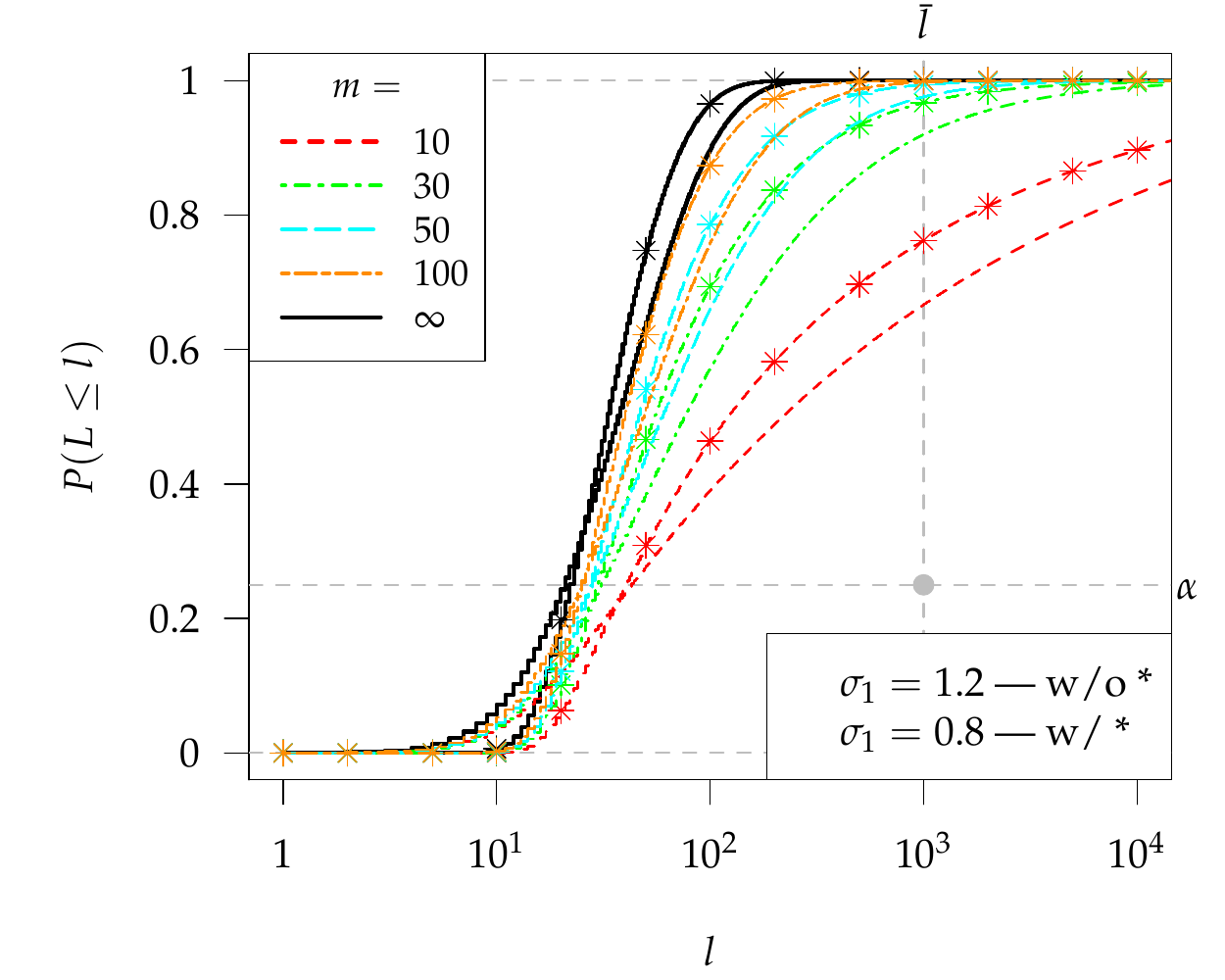}
	\end{tabular}
	\caption{Unconditional OOC RL CDFs for EWMA ($\lambda=0.1$) $S^2$ ($n=5$), phase I size $m$,
		$P_\text{IC}(L\le 10^3) = 0.25$.}\label{fig:05}
\end{figure}

Widening the limits allows poor false alarm levels to be dealt with. However, the behavior in the OOC case
has deteriorated. Using the limits from $P_\text{IC}(L\le 1\,000) = 0.25$, we show
the unconditional CDFs for selected OOC cases
($\sigma_1 \in \{0.8, 1.2\}$)
in Figure~\ref{fig:05}.
Note the poor detection behavior for smaller values of $m$. For $m < 50$, it is possible that
the variance change will remain undetected over the entire planned monitoring time span ($\bar{l} = 1\,000$ observations). It is even worse for the two-sided case.
Based on the profiles in Figure~\ref{fig:05}, we would recommend phase I sizes of at least 100.
For more details, we refer the reader to the next section.

\begin{figure}[hbt] 
\renewcommand{\tabcolsep}{-.7ex}
\begin{tabular}{cc}
  \footnotesize upper & \footnotesize two-sided \\[-2ex]
  \includegraphics[width=.52\textwidth]{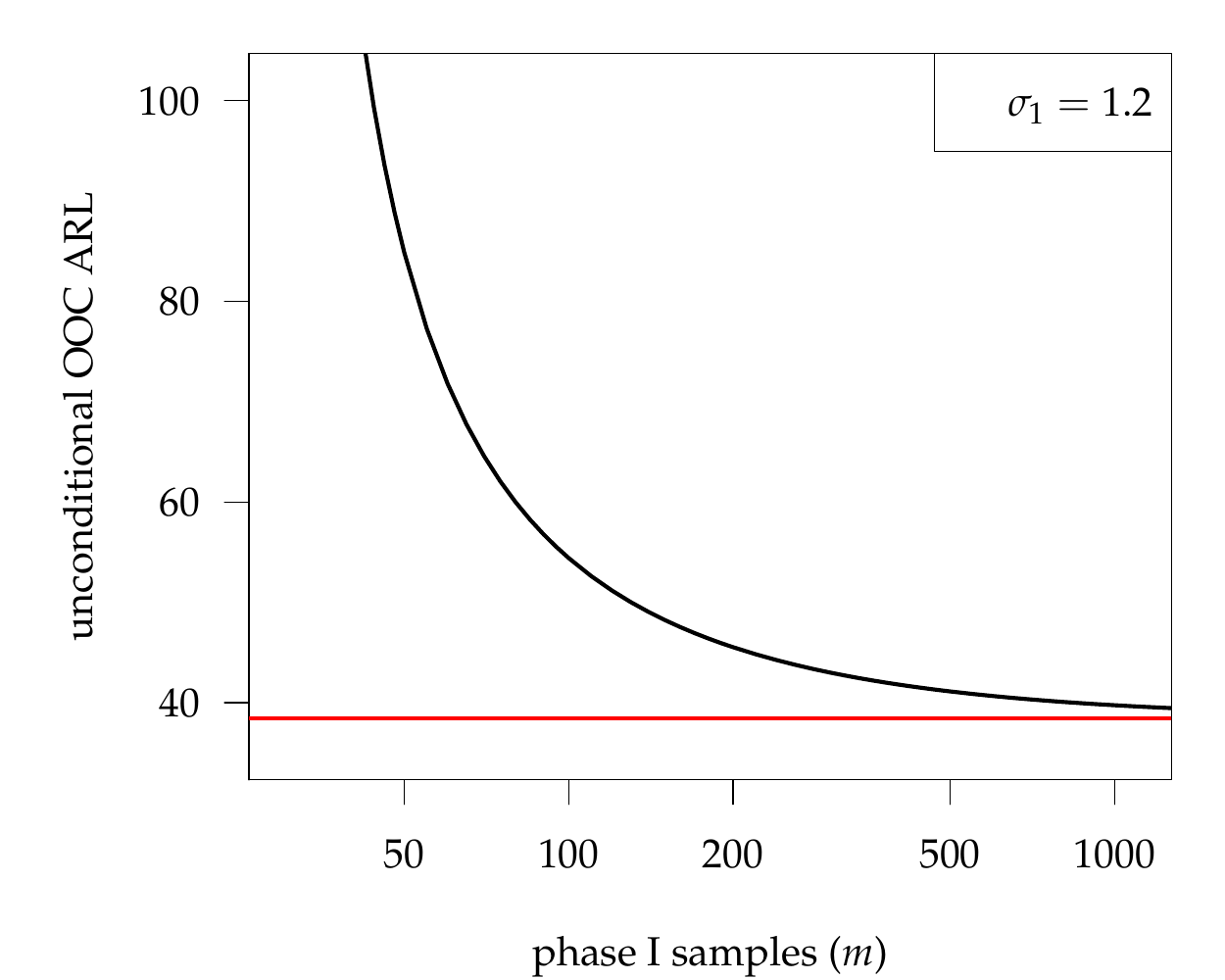}	& 
  \includegraphics[width=.52\textwidth]{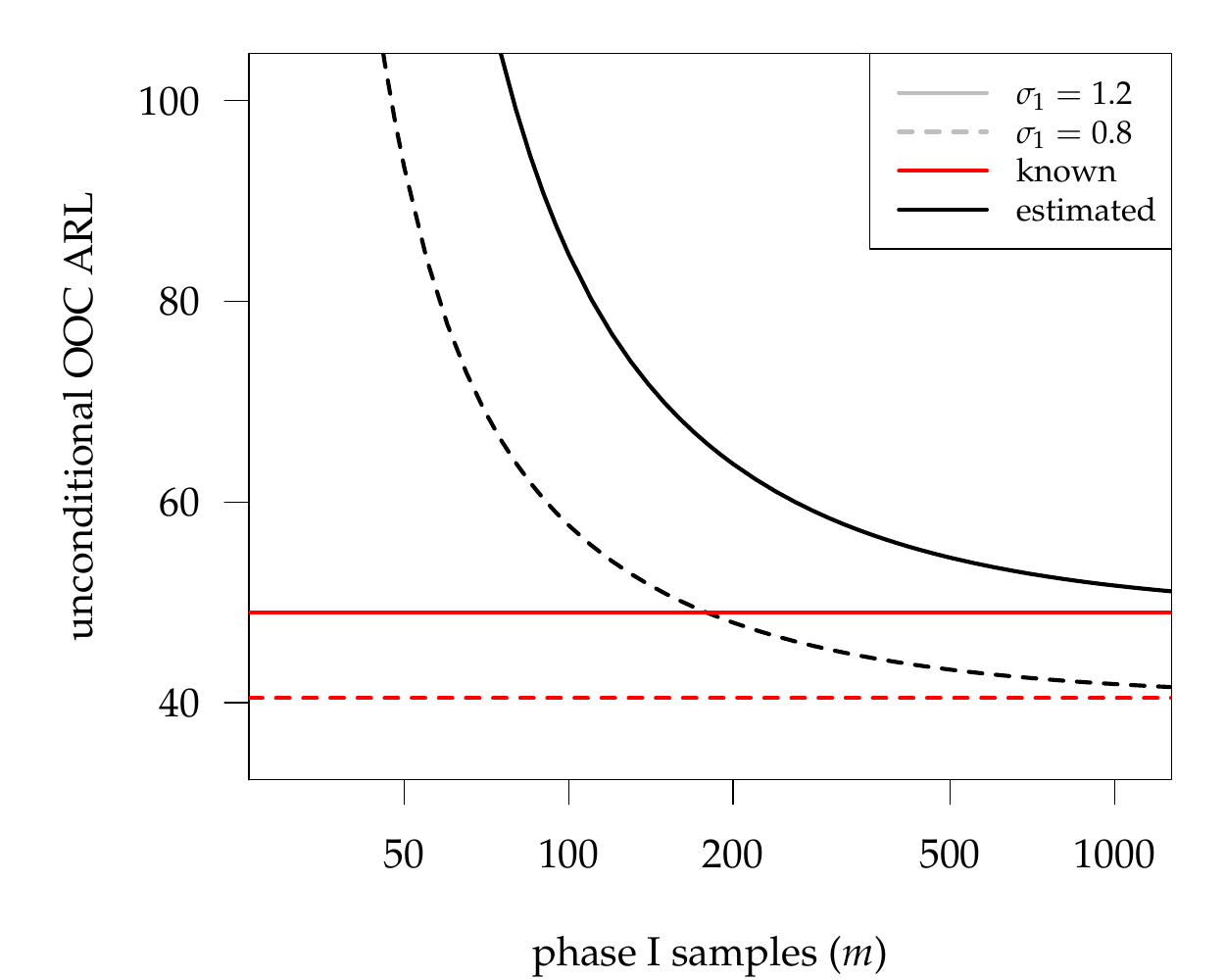}
\end{tabular}
\caption{Unconditional OOC ARLs for EWMA ($\lambda=0.1$) $S^2$ ($n=5$) vs. phase I size $m$,
$P_\text{IC}(L\le 10^3) = 0.25$.}\label{fig:06}
\end{figure}

In order to provide some more familiar representations and at least get an idea of the detection speed, we add
some unconditional ARL values to the OOC case in Figure~\ref{fig:06}.
The differences in the benchmark case are considerably large for $m < 100$ and become negligible only for $m > 200$. Hence, there is an
obvious price to pay if we account for the phase I estimation uncertainty when calibrating the chart.
Because the false alarm behavior is really important for practical control charting in industry,
the calibration strategy utilizing $P(L \le \bar{l}) = \alpha$ seems to be passable despite these side effects.
Note that the even more conservative approach of guaranteeing a minimum conditional IC ARL
yields substantially larger unconditional OOC ARL results.

\section{Sensitivity and competition} \label{sec:comp}

To reconcile the common IC ARL user to this method, we investigate the selection of the monitoring horizon $\bar l$ and false alarm probability $\alpha$
for a given IC ARL of, for example, 500 and its impact on the actual adjustment of the control limits accounting for
the estimation uncertainty. To begin with, we set $\alpha = 0.5$ to
\begin{figure}[hbt]
\renewcommand{\tabcolsep}{-.7ex}
\begin{tabular}{cc}	
  \footnotesize upper & \footnotesize two-sided \\[-2ex]
  \includegraphics[width=.52\textwidth]{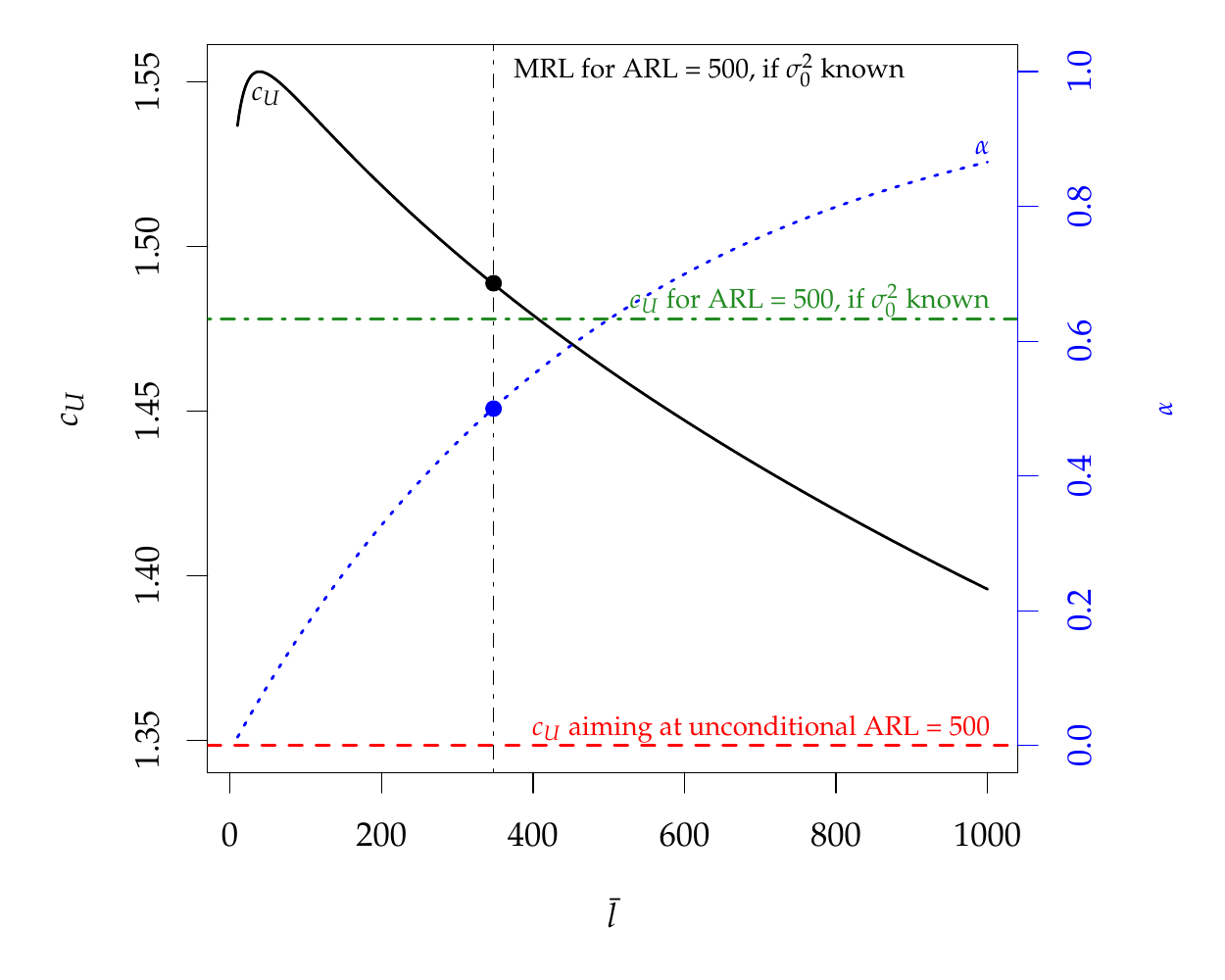} &		
  \includegraphics[width=.52\textwidth]{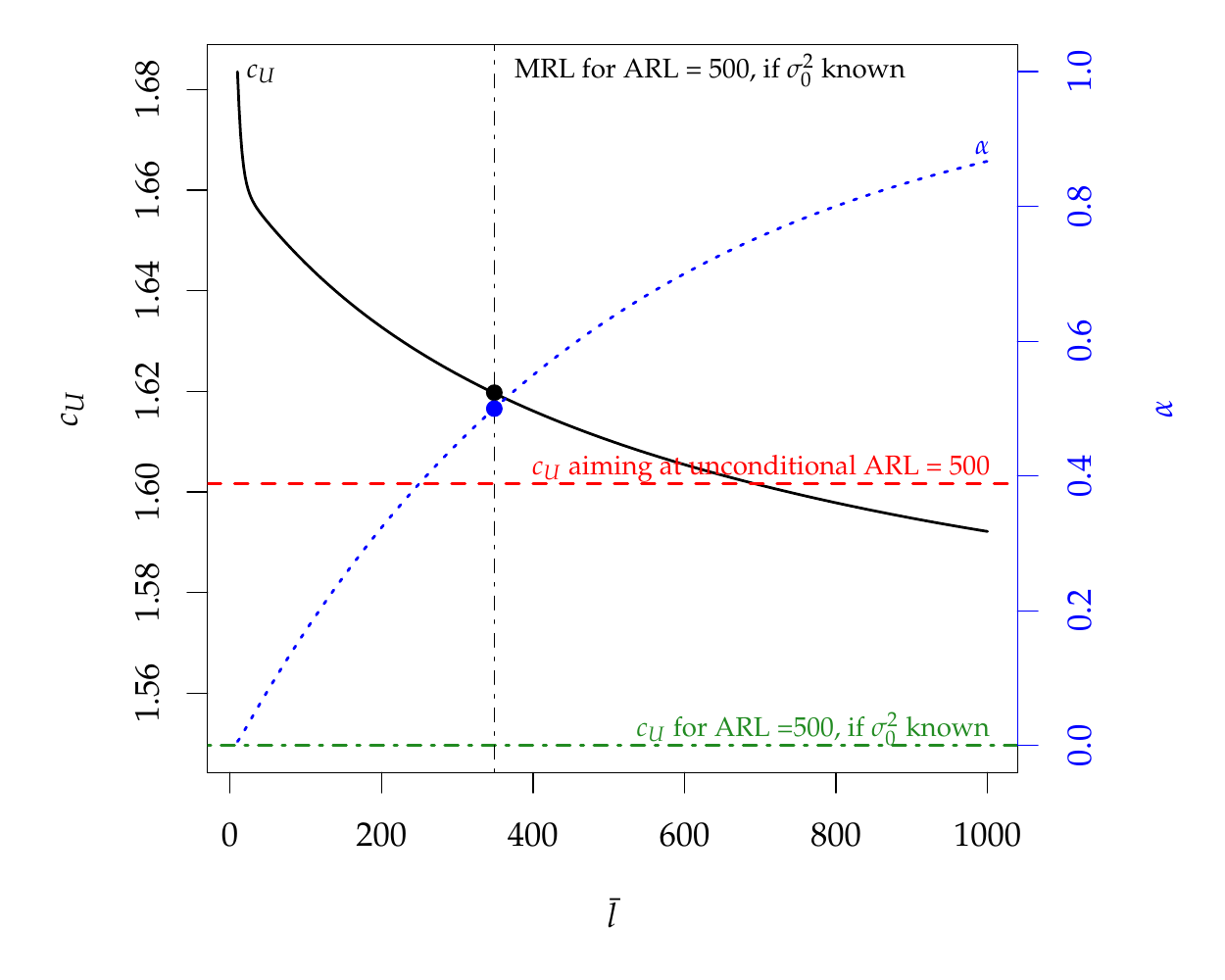}		
\end{tabular}
\caption{Choice of $(\bar l, \alpha)$ within $\alpha = P(L\le \bar l)$ and $E(L) = 500$ (all for known $\sigma_0^2$)
and its impact on the $c_U$ modification to secure $\alpha = P(L\le \bar l)$ in the case of unknown $\sigma_0^2$, which will be estimated
with a $m=50 \times n=5$ phase I sample.}\label{fig:07}
\end{figure}
ensure that the IC median run length (MRL) 348 (349 in the two-sided case) is achieved. From Figure~\ref{fig:07}(a) , we conclude that
focusing on the unconditional ARL yields the smallest $c_U$, followed by simply utilizing the $c_U$ value for known $\sigma_0^2$ and,
finally, the unconditional MRL (median RL) design. Obviously, downsizing the upper limit seems to be counter-intuitive and results in more false alarms
than intended. The slight increase of $c_U$ from the known $\sigma_0^2$ case to the MRL- conserving approach offers a cautious 
and effective way of dealing with the estimation uncertainty. By changing $\alpha$ (or $\bar l$), we can see that for all $\alpha < 0.56$, the
modified $c_U$ is larger than for known $\sigma_0^2$.
In addition, decreasing $\alpha$ (and $\bar{l}$, accordingly) increases $c_U$ further (except for very small $\alpha$).
Of course, proper choices of $\alpha$ are 0.5 or smaller.
In the two-sided case, all designs securing some unconditional measure widen the original limits. In Figure~\ref{fig:07}(b),
we plot only the upper value $c_U$. For $\alpha < 0.75$, the unconditional RL quantiles induce wider limits than the unconditional
ARL design. In summary, deciding on a reasonable combination  $(\bar l, \alpha)$ provides plausible and effective limit adjustments
that can overcome the estimation uncertainty distortions.

Next, we wish to compare the detection behaviors of various values of the smoothing constant $\lambda \in \{0.05, 0.1, 0.2, 0.3\}$.
In all cases, we calibrate the schemes to ensure that $P_\text{IC}(L\le 10^3)=0.25$.
Again, we consider samples of size $n = 5$ and a phase I study of size $m = 50$.
In Table~\ref{Tab:02} and Table~\ref{Tab:03}, we provide some (unconditional) ARL values
\begin{table}[hbt]
\centering
\begin{tabular}{cccccc} \toprule
  $\lambda$	& 0.05 & 0.1 & 0.2 & 0.3 & 1$_\text{\tiny Shewhart}$ \\ \toprule
  \multicolumn{6}{c}{known $\sigma_0^2$} \\ \midrule
  $c_u$ & 1.3995 & 1.6453 & 2.0690 & 2.4653 & 5.3026 \\ \midrule
  $E_1(L)$     & 3\,444 & 3\,461 & 3\,470 & 3\,473 & 3\,476 \\
  $E_{1.2}(L)$ & 32.9 & 38.4 & 55.9 & 76.1 & 188.8 \\
  $E_{1.5}(L)$ & 8.75 & 8.05 & 8.24 & 9.11 & 19.5 \\ \midrule
  \multicolumn{6}{c}{phase I with $m=50$} \\ \midrule
  $c_u$ & 1.4680 & 1.7198 & 2.1538 & 2.5596 & 5.4654 \\ \midrule
  $E_1(L)$ & $> 5\times 10^9$ & $>8\times 10^5$ & 47\,128 & 21\,477 & 8\,091 \\
  $E_{1.2}(L)$ & 70.4 & 84.8 & 119.2 & 151.8 & 293.4 \\
  $E_{1.5}(L)$ & 10.6 & 9.52 & 9.79 & 11.0 & 24.0 \\ \bottomrule
\end{tabular}
\caption{Unconditional ARL results for various values of $\lambda$ and $P_\text{IC}(L\le 10^3)=0.25$, upper chart.}\label{Tab:02}
\end{table}
for the upper and the two-sided designs, respectively. 
To judge phase I‘s influence on the uncertainty, we compare the unconditional ARL numbers with the
initial ones for a known IC variance. Interestingly, the
new IC ARL results are very large but decline with increasing $\lambda$.
Similar patterns can be observed in the OOC case, where for $\sigma = 1.2$, the ARL numbers are
doubled. For the medium size increase, that is, $\sigma = 1.5$, the change is much smaller.
Note that the order between the different EWMA designs remains stable, that is, $\lambda = 0.05$ is the best
for $\sigma = 1.2$, while $\lambda = 0.1$ works best with $\sigma = 1.5$.
The Shewhart $S^2$ chart ARL results are added, which are considerably larger than all EWMA ones.

Turning to the two-sided designs, we observe some slight differences. Most notably, the IC ARL values do not explode.
\begin{table}[hbt]
\centering
\begin{tabular}{cccccc} \toprule
  $\lambda$	& 0.05 & 0.1 & 0.2 & 0.3 & 1$_\text{\tiny Shewhart}$ \\ \toprule
  \multicolumn{6}{c}{known $\sigma_0^2$} \\ \toprule
  $c_l, c_u$ & 0.6825, 1.4377 & 0.5610, 1.7051 & 0.4146, 2.1721 & 0.3200, 2.6159 & 0.0112, 6.3542 \\ \midrule
  $E_{0.5}(L)$ & 11.3 & 8.91 & 7.35 & 6.99 & 264 \\
  $E_{0.8}(L)$ & 35.3 & 40.5 & 68.0 & 113 & 1\,671 \\
  $E_1(L)$ & 3\,425 & 3\,453 & 3\,462 & 3\,467 & 3\,465 \\
  $E_{1.2}(L)$ & 38.9 & 49.0 & 81.1 & 121 & 639.5 \\
  $E_{1.5}(L)$ & 9.64 & 8.96 & 9.53 & 11.1 & 42.6 \\ \midrule
  \multicolumn{6}{c}{phase I with $m=50$} \\ \midrule
  $c_l, c_u$ & 0.6377, 1.5488 & 0.5287, 1.8249 & 0.3947, 2.3077 & 0.3067, 2.7669 & 0.0111, 6.6824 \\ \midrule
  $E_{0.5}(L)$ & 13.5 & 10.0 & 8.03 & 7.60 & 277 \\  
  $E_{0.8}(L)$ & 75.4 & 93.3 & 151 & 222 & 1\,752 \\  
  $E_1(L)$ & 11\,240 & 6\,803 & 4\,961 & 4\,386 & 3\,500 \\
  $E_{1.2}(L)$ & 140 & 173 & 251 & 328 & 1\,094 \\
  $E_{1.5}(L)$ & 12.9 & 11.5 & 12.4 & 14.8 & 62.6 \\ \bottomrule
\end{tabular}
\caption{Unconditional ARL results for various values of $\lambda$ and $P_\text{IC}(L\le 10^3)=0.25$, two-sided chart.}\label{Tab:03}
\end{table}
Again, the OOC ARL results are tripled ($1.2$) and doubled ($0.8$) for small changes,
while the increases are quite small for larger variance changes ($0.5, 1.5$).
For the simple Shewhart chart, the unconditional ARL values nearly coincide with
their known $\sigma_0^2$ counterparts (increasing $\sigma$ only).
For the upper and two-sided designs, the patterns in the detection ranking
remain stable.
For small shifts, for example, the EWMA chart with $\lambda = 0.05$  exhibits the best detection
performance for known and unknown IC variances.
It should be stated that the Shewhart ARL performance is much worse than the EWMA ARL performance for the changes considered
here.

\begin{figure}[hbt] 
\renewcommand{\tabcolsep}{-.7ex}
\begin{tabular}{cc}
  \footnotesize upper & \footnotesize two-sided \\[-2ex]
  \includegraphics[width=.5\textwidth]{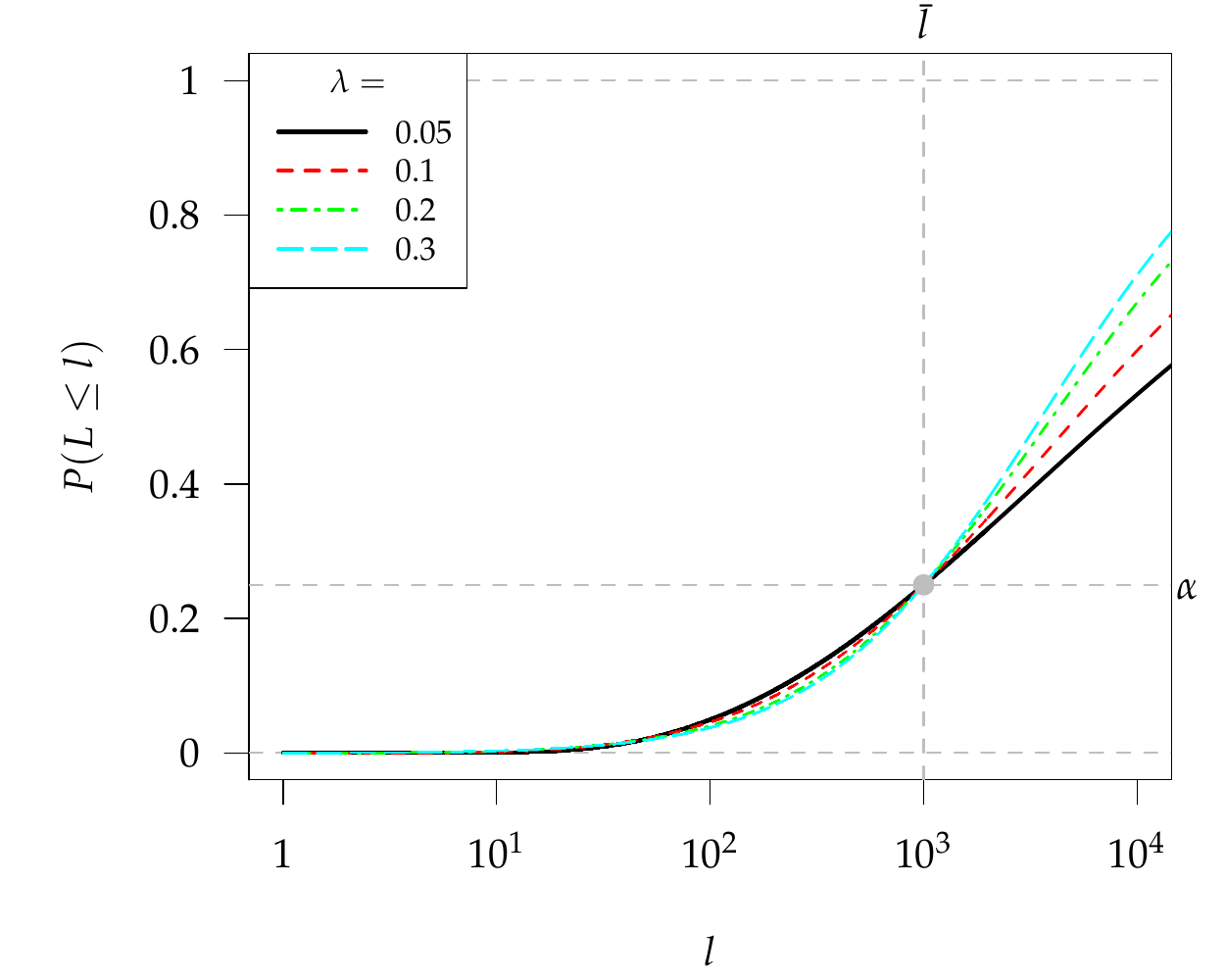} & 
  \includegraphics[width=.5\textwidth]{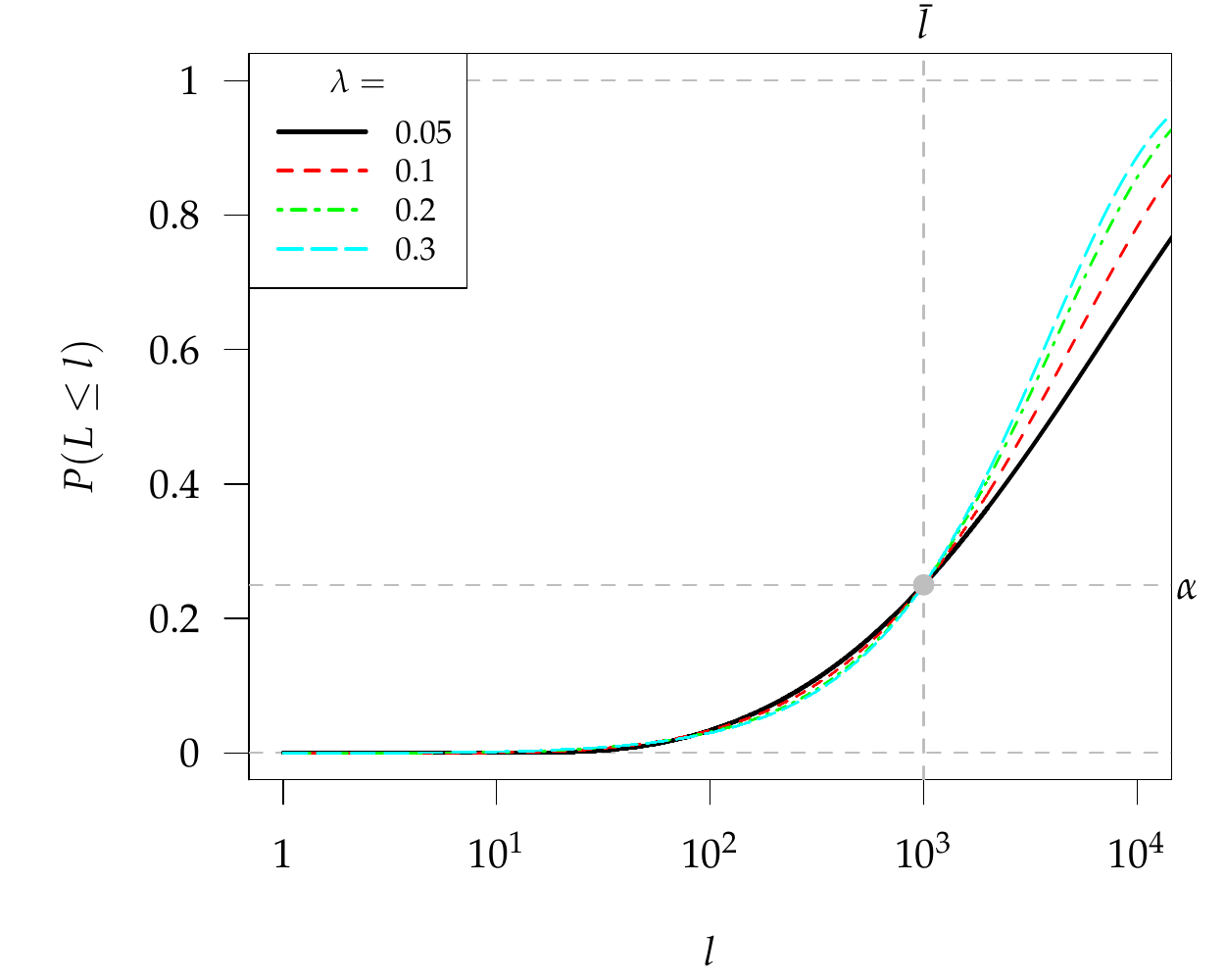}
\end{tabular}
\caption{IC CDFs of the RL $L$, $P_\text{IC}(L\le 10^3)=0.25$, EWMA (various $\lambda$) $S^2$ ($n=5$), $m=50$ phase I samples.}\label{fig:08}	
\end{figure}

Following the focus of this paper, we now examine the CDF profiles. Beginning with the IC versions for both designs in Figure~\ref{fig:08},
we conclude that for $l \le \bar l = 1\,000$, the profiles look very similar. The $\lambda = 0.05$ line lies above all the others for these $l$,
which changes for $l > \bar l$, where it features the lowest values. All other curves follow according their $\lambda$ values, that is, the
larger the $\lambda$, the lower the $l \le \bar l$ and the higher the $l > \bar l$.
Comparing the results for the upper and two-sided EWMA designs, we observe much steeper developments for the latter ones.
In summary, we conclude that for the interesting part, namely, $l \le \bar l$, the IC behavior of $P(L\le l)$
for all considered $\lambda$ values and for both design types does not really differ.

\begin{figure}[hbt] 
\renewcommand{\tabcolsep}{-.7ex}
\begin{tabular}{cc}
  \footnotesize $\sigma_1=1.2$ & \footnotesize $\sigma_1=1.5$ \\[-2ex]
  \includegraphics[width=.5\textwidth]{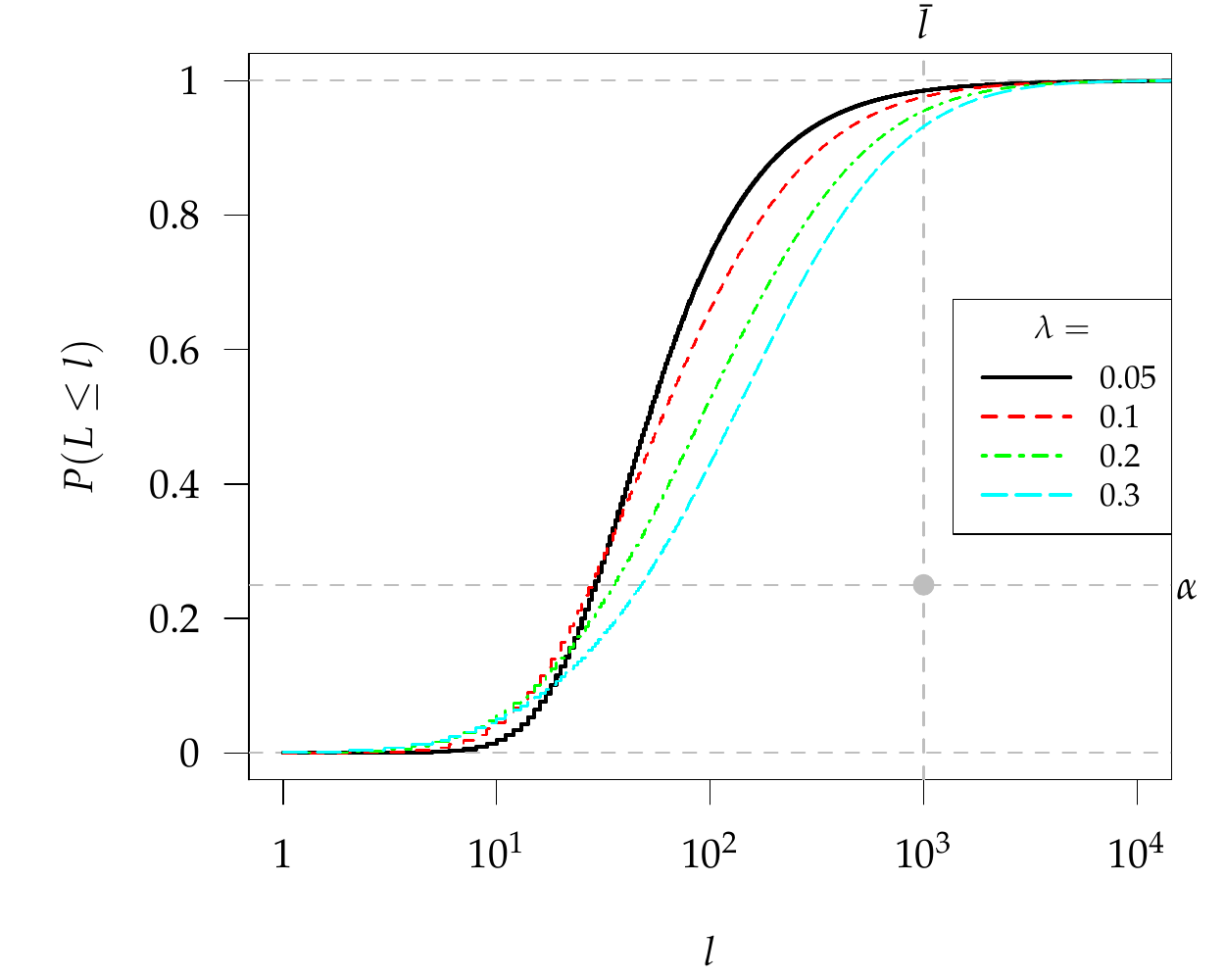} &
  \includegraphics[width=.5\textwidth]{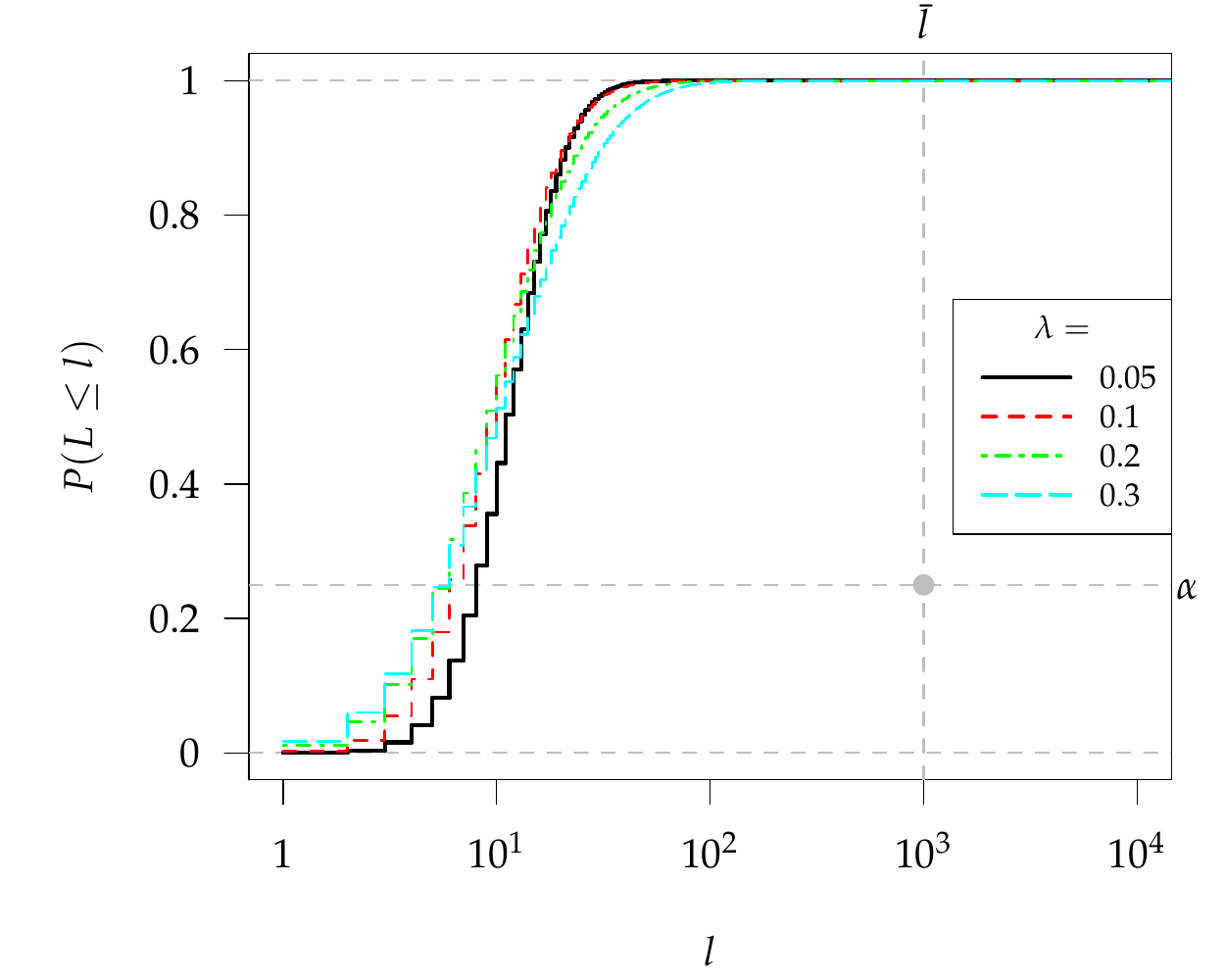} \\
  \footnotesize $\sigma_1=0.8$ & \footnotesize $\sigma_1=0.5$ \\[-2ex]
  \includegraphics[width=.5\textwidth]{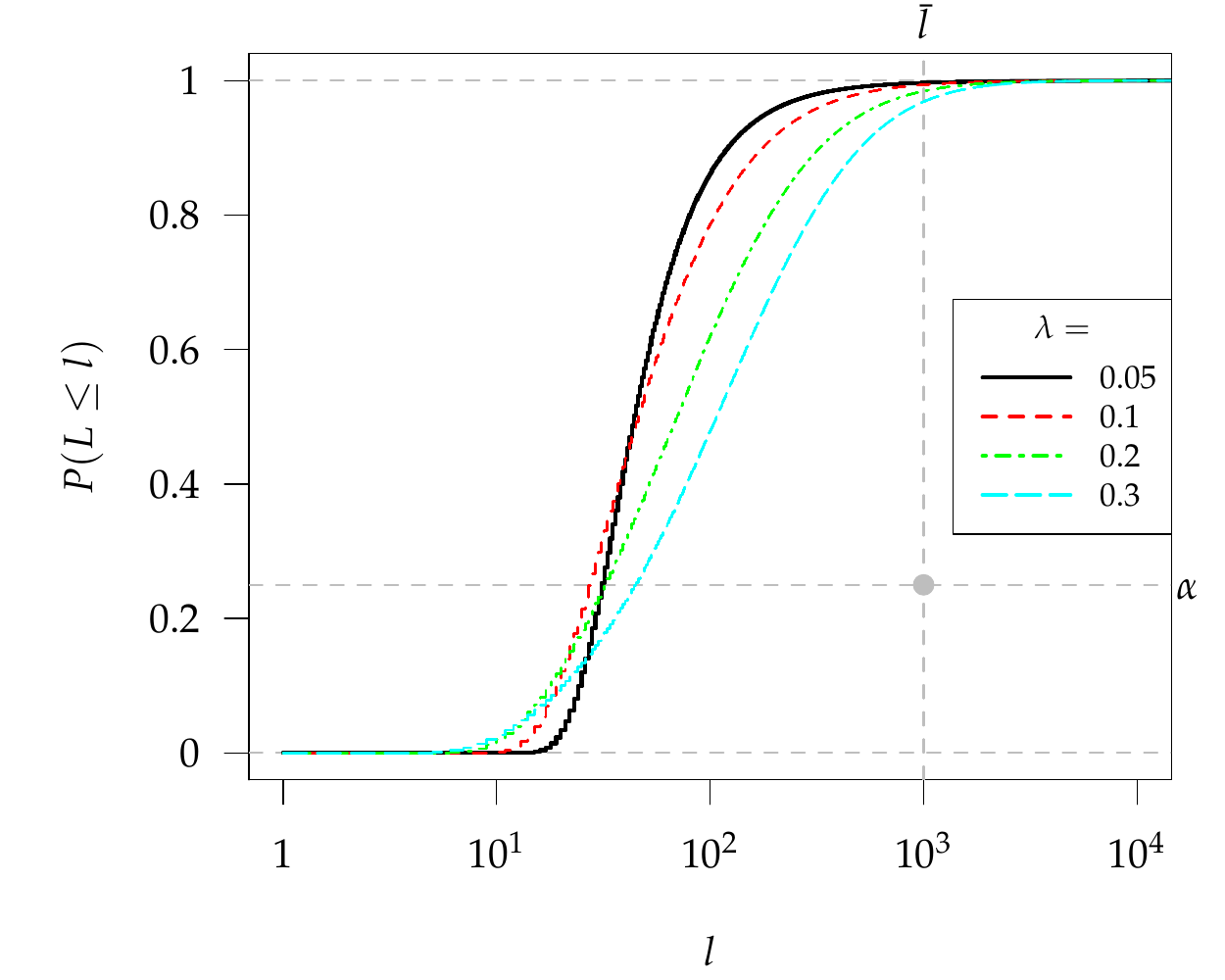} &
  \includegraphics[width=.5\textwidth]{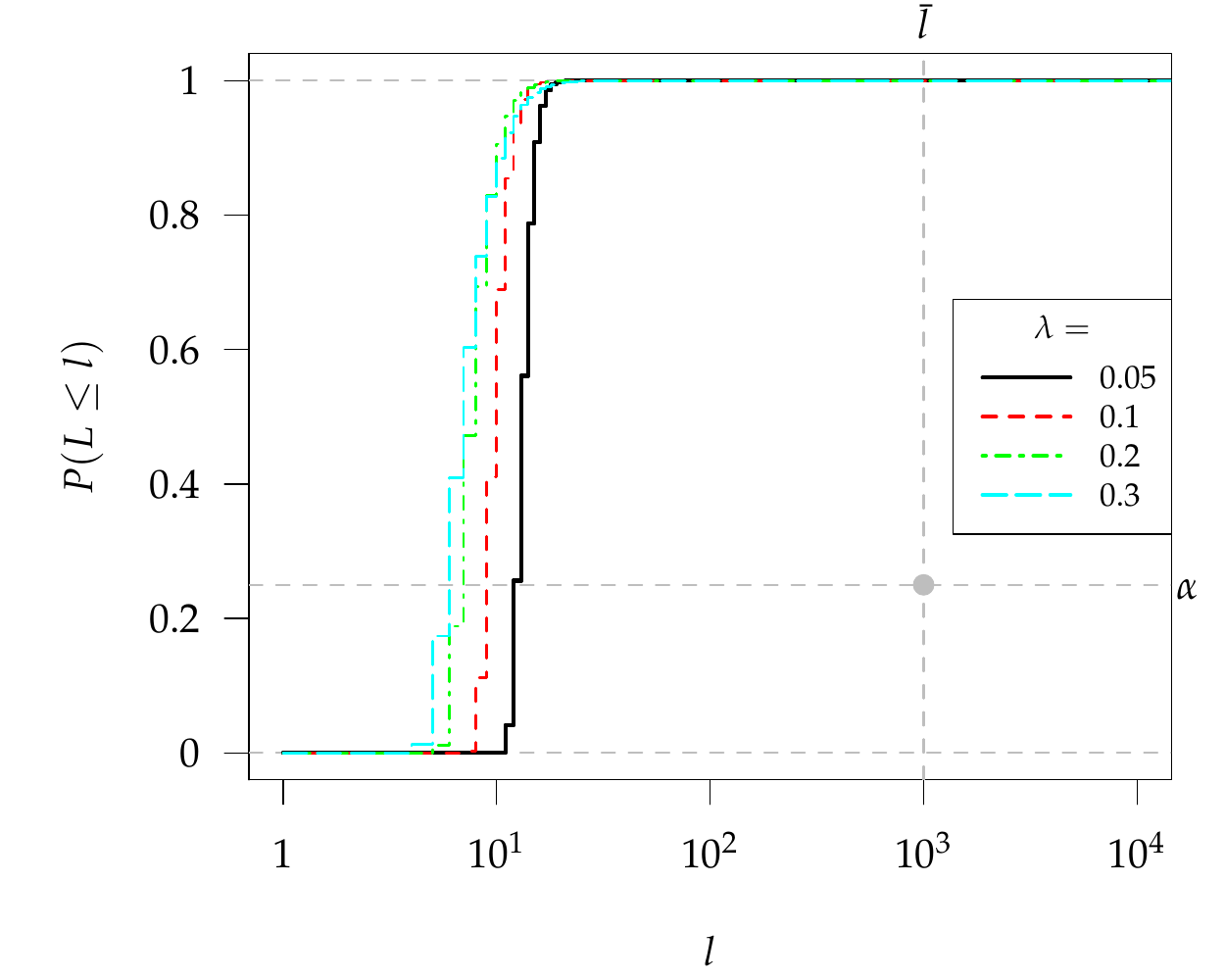}
\end{tabular}
\caption{OOC CDFs of the RL $L$, $P_\text{IC}(L\le 10^3)=0.25$, two-sided EWMA (various $\lambda$) $S^2$ ($n=5$), $m=50$ phase I samples.}\label{fig:10}	
\end{figure}

Turning to the OOC case, we start with the upper EWMA chart and two different possible new $\sigma$ values,
\begin{figure}[hbt] 
\renewcommand{\tabcolsep}{-.7ex}
\begin{tabular}{cc}
  \footnotesize $\sigma_1=1.2$ & \footnotesize $\sigma_1=1.5$ \\[-2ex]
  \includegraphics[width=.5\textwidth]{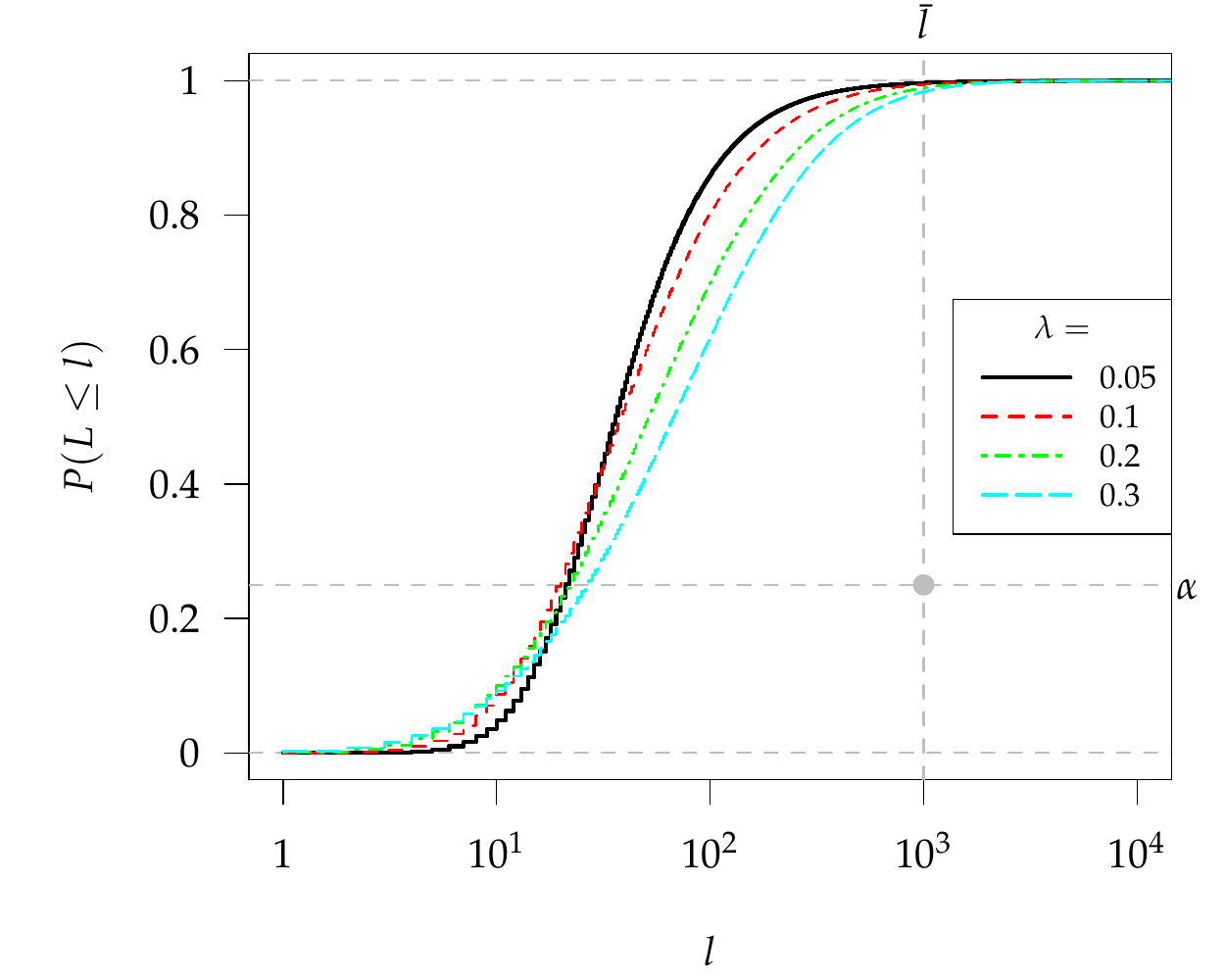} & 
  \includegraphics[width=.5\textwidth]{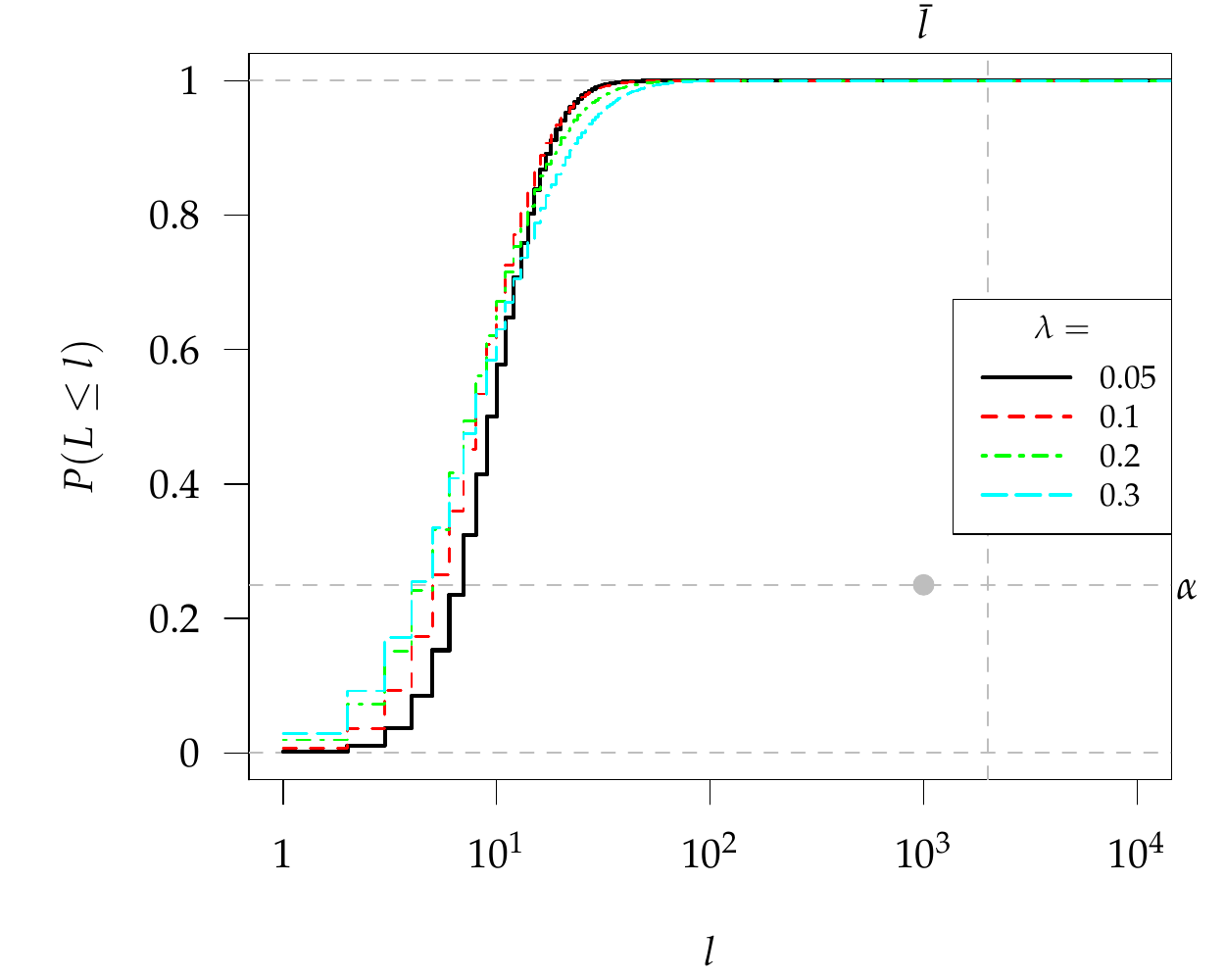}
\end{tabular}
\caption{OOC CDFs of the RL $L$, $P_\text{IC}(L\le 10^3)=0.25$, upper EWMA (various $\lambda$) $S^2$ ($n=5$), $m=50$ phase I samples.}\label{fig:09}	
\end{figure}
namely, the small change $\sigma_1 = 1.2$ and the medium one $\sigma_1 = 1.5$. Examining Figure~\ref{fig:09}, we observe the following stylized facts.
The larger change is detected by $l \le 100$ with probability one, while for $\sigma_1 = 1.2$, we need the whole time span, that is, $l \le 1\,000$.
Recall the corresponding expected values in Table~\ref{Tab:02}, which are roughly 10 for $\sigma_1 = 1.5$ for all EWMA designs,
while for $\sigma_1 = 1.2$, they range from 70 for $\lambda=0.05$ to 150 for the largest $\lambda < 1$.
Moreover, the order between these $\lambda$ values defined by
their ARL values is reflected by the $P(L\le l)$ profiles. Similar patterns can be recognized for the two-sided designs in Figure~\ref{fig:10},
where we included the results for decreased variances. Not surprisingly, the detection performance for $\sigma_1 \in\{1.2,1.5\}$
is weaker compared to the upper chart profiles in Figure~\ref{fig:09}. However, detecting decreases of the same relative order proceeds more
quickly.

\begin{figure}[hbt] 
\renewcommand{\tabcolsep}{-.7ex}
\begin{tabular}{cc}
  \scriptsize (a) lower limit & \scriptsize (b) upper limit \\[-1ex]	
  \includegraphics[width=.5\textwidth]{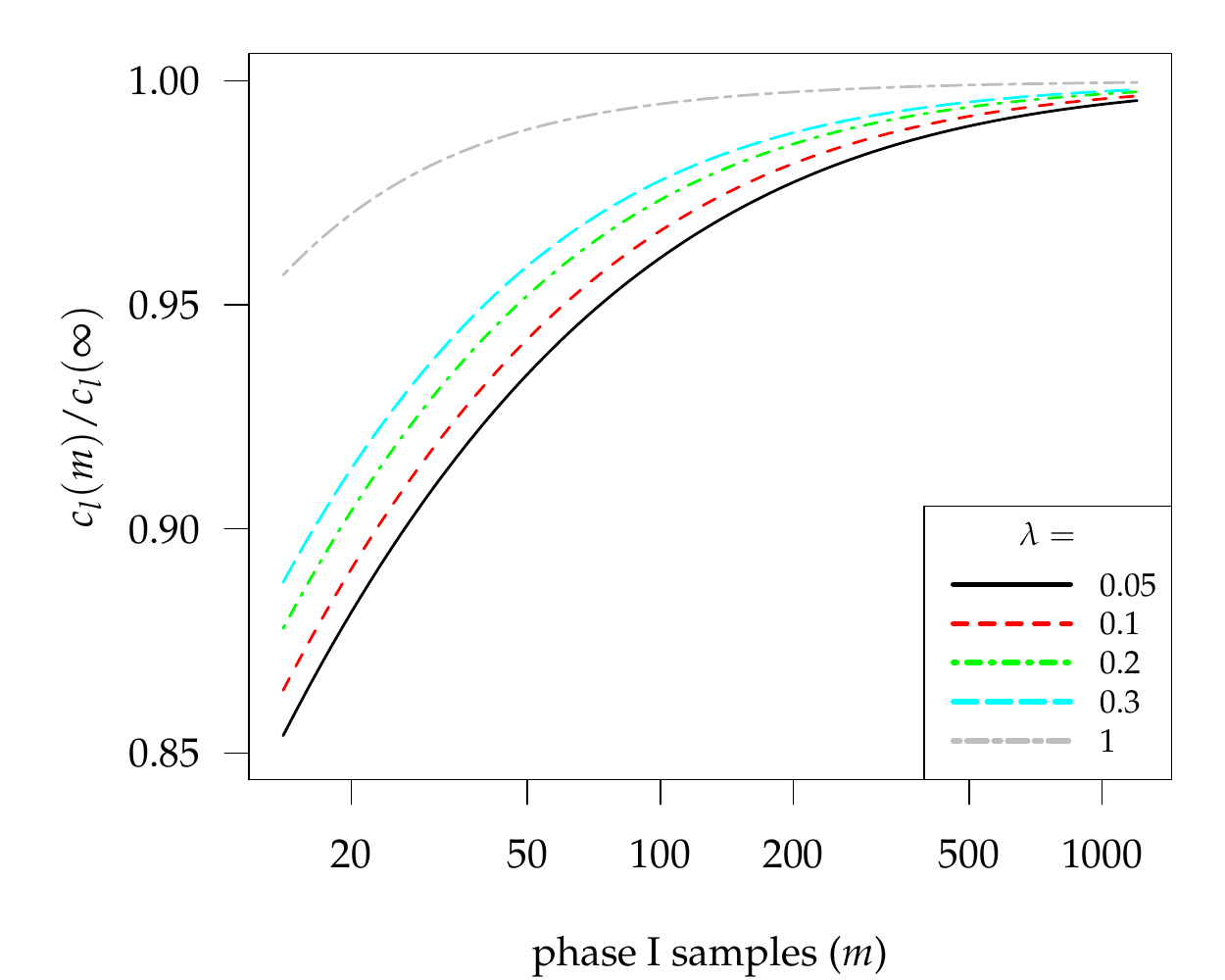} &
  \includegraphics[width=.5\textwidth]{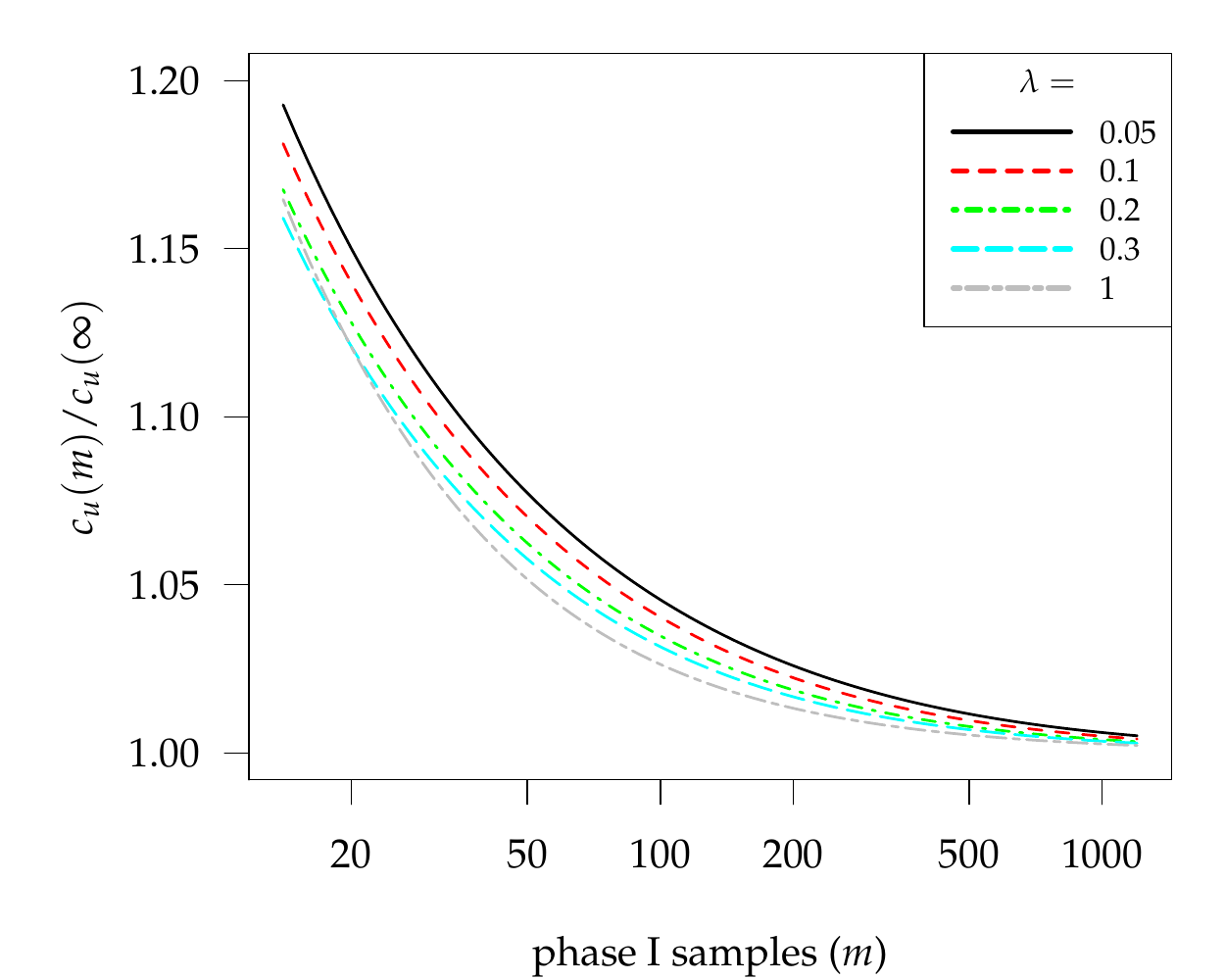} \\ 
\end{tabular}
\caption{Ratios of the modified control limits to the original ones (known IC variance),
$P_\text{IC}(L\le 10^3)=0.25$, two-sided EWMA (various $\lambda$) $S^2$ ($n=5$), phase I size $m=15, 16, \ldots, 1200$.}\label{fig:12}	
\end{figure}

After some first glimpses of the impact of the phase 1 sample size, namely, $m$, on the magnitude of
the limit modification in Figure~\ref{fig:03}, the resulting unconditional OOC ARLs in Figure~\ref{fig:06}
and the snapshots in Tables~\ref{Tab:02} and \ref{Tab:03},
some additional details will be provided here to develop recommendations regarding some lower bound
for $m$ and the choice of the smoothing constant $\lambda$. We start with the upper design
and look at our selection of EWMA smoothing constants $\lambda \in \{0.05, 0.1, 0.2, 0.3\}$.
In the following Figure~\ref{fig:11}, the $c_u$ vs. phase I size $m$ profiles are provided in two
ways. First, the raw $c_u$ limits are presented, demonstrating the typical behavior of
decreasing values if $\lambda \downarrow$ or $m \uparrow$. Note that the change from $\lambda = 0.3$
\begin{figure}[hbt] 
\renewcommand{\tabcolsep}{-.7ex}
\begin{tabular}{cc}
  \scriptsize (a) raw size & \scriptsize (b) relative change to known parameter case \\[-1ex]	
  \includegraphics[width=.5\textwidth]{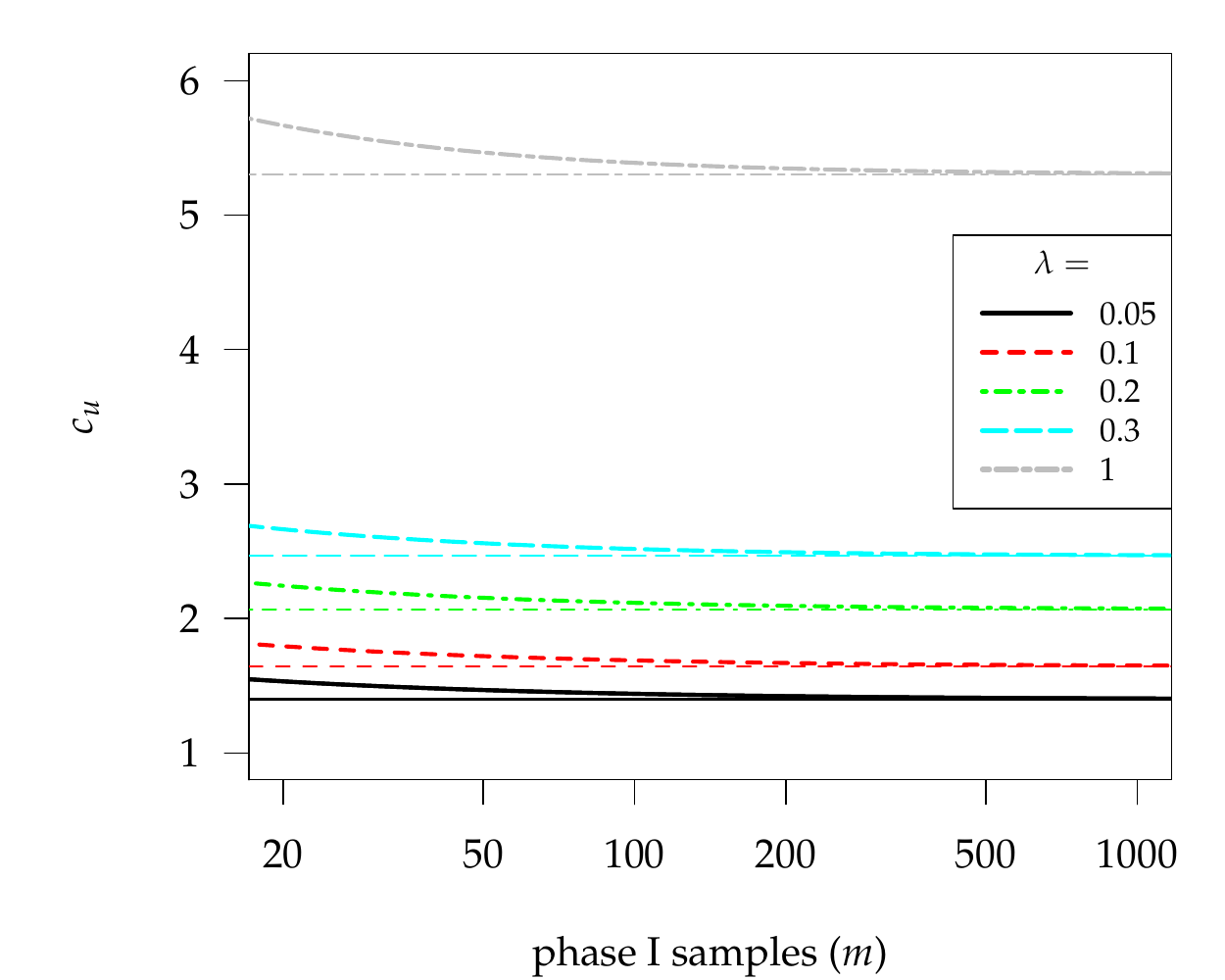} &
  \includegraphics[width=.5\textwidth]{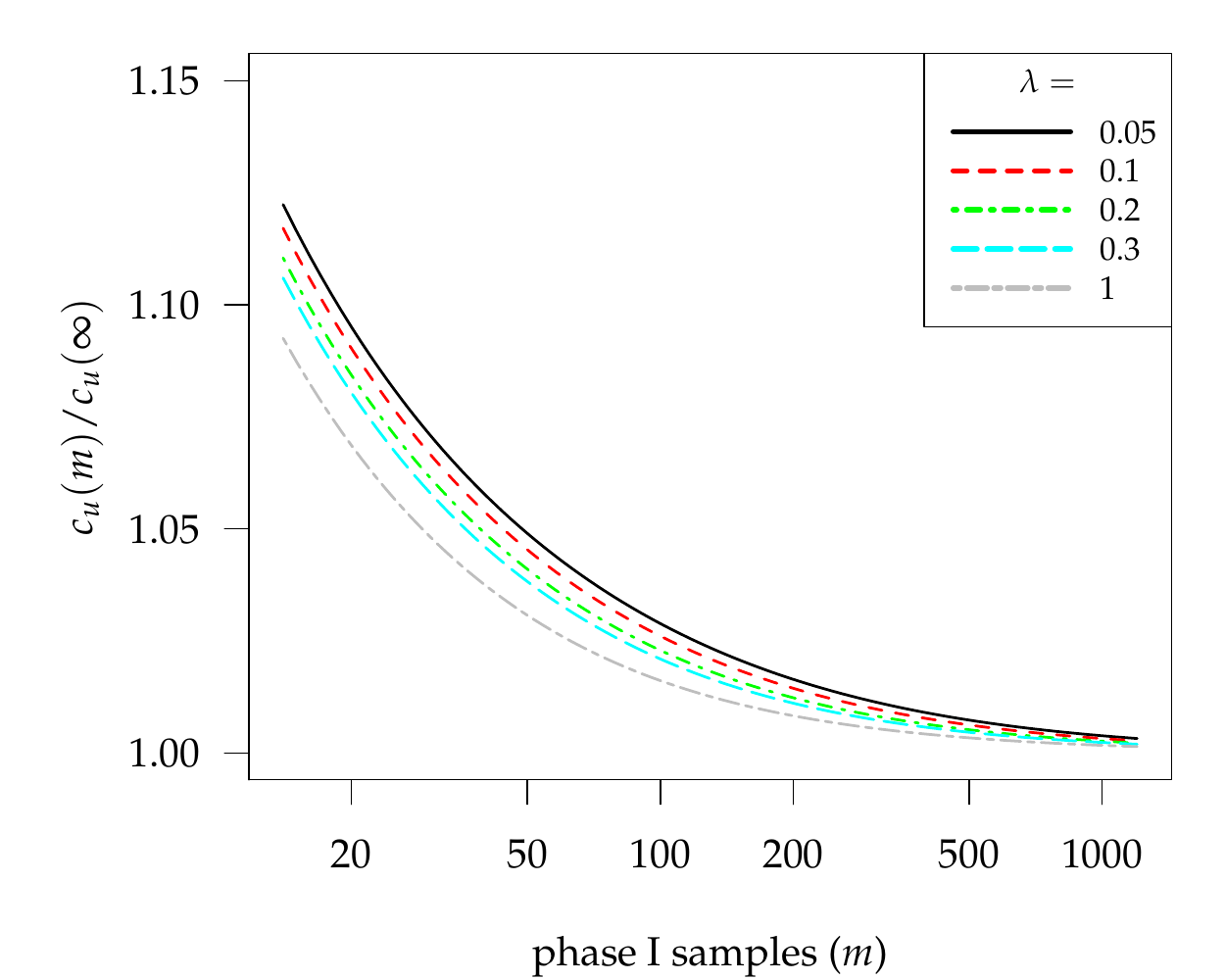} \\ 
\end{tabular}
\caption{Modified $c_u$ needed to achieve $P_\text{IC}(L\le 10^3)=0.25$, upper EWMA (various $\lambda$) $S^2$ ($n=5$), phase I size $m=15, 16, \ldots, 1200$.}\label{fig:11}	
\end{figure}
to the Shewhart case ($\lambda = 1$) is really pronounced. However, the amount of widening in the
control chart's continuation region done
to cope with estimating the IC value of the variance is not large.
Compared to the known parameter case, illustrated in Figure~\ref{fig:11}(b), we must increase the
original $c_u$ by 5 to 10\% (along the $\lambda$ range) for small $m < 50$ and by less than 3\% for $m > 100$.
The relative amount of change decreases with increasing $\lambda$.
Similar results can be observed for the two-sided case.
Next, we consider just the relative changes plotted in Figure~\ref{fig:12}.
Except for the lower limit in the Shewhart chart, which is driven by the small sample size $n = 5$ creating
difficulties while detecting variance decreases and is typically very close to zero,
the profiles do not really differ from those for the one-sided case.
Not surprisingly, the adjustment needed is larger than for the one-sided design.
Overall, the widening of the control chart limits is about 10\% or smaller for $m \ge 30$.
Even some crude rule of thumb for selected values of $m$ could be derived, such as widen the limits by
10, 8, 5, 3, 2 and 1\% for $m = $ 20, 30, 50, 100, 200 and 400  for the phase I samples, respectively.
In conclusion, utilizing the $P(L\le \bar l)=\alpha$ design yields moderate changes to the control
chart limits. To identify a minimum $m$ rule or a $\lambda$ guideline, we now consider 
the unconditional OOC ARL for two magnitudes of change.

\begin{figure}[hbt] 
\renewcommand{\tabcolsep}{-.7ex}
\begin{tabular}{cc}
  \scriptsize (a) $\sigma_1 = 1.2$ & \scriptsize (b) $\sigma_1 = 1.5$ \\[-1ex]	
  \includegraphics[width=.5\textwidth]{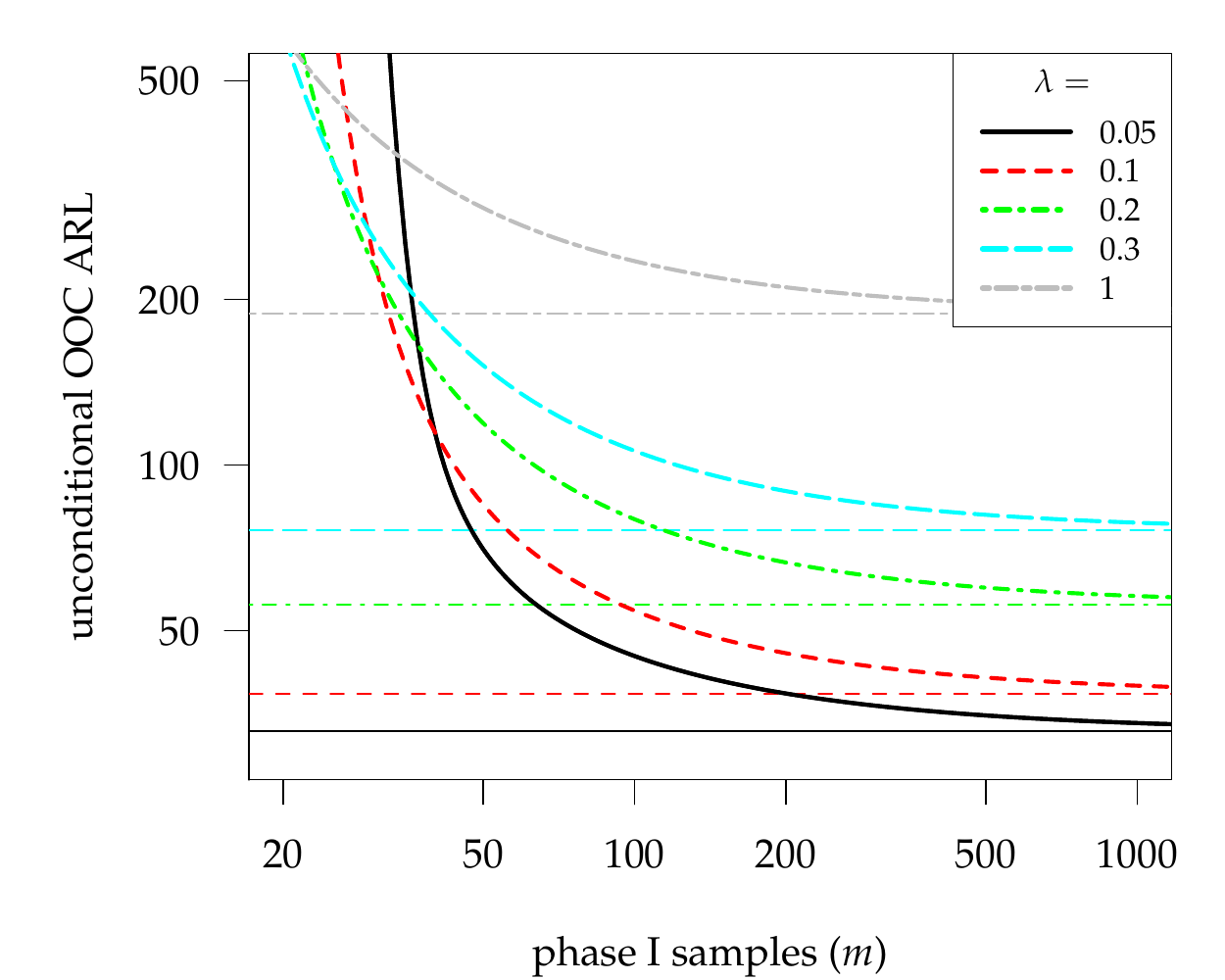} &
  \includegraphics[width=.5\textwidth]{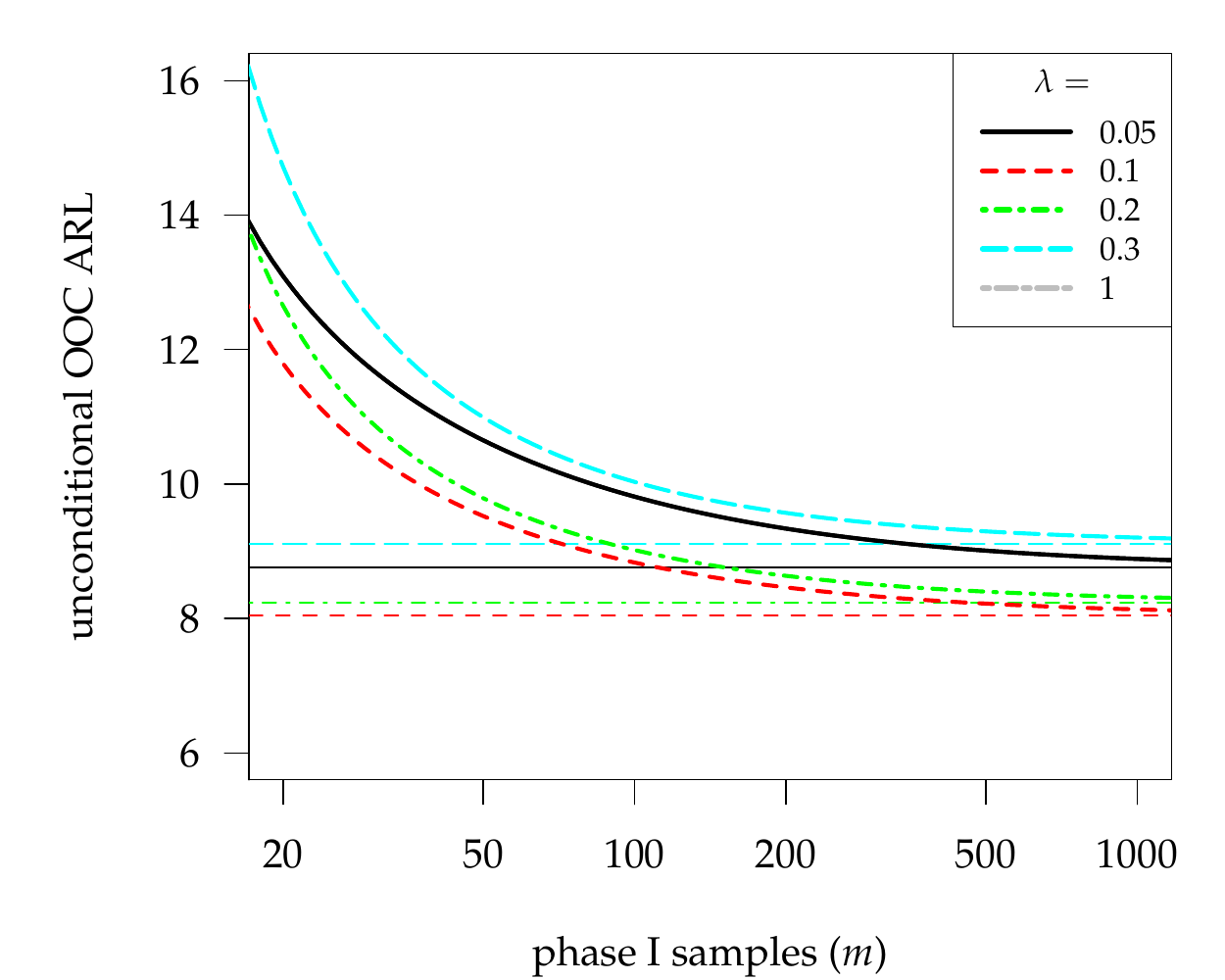} \\  
\end{tabular}
\caption{Unconditional OOC ARLs of the upper EWMA (various $\lambda$) $S^2$ ($n=5$) charts,
$P_\text{IC}(L\le 10^3) = 0.25$, phase I size $m=15, 16, \ldots, 1200$.}\label{fig:13}	
\end{figure}

\begin{figure}[hbt] 
\renewcommand{\tabcolsep}{-.7ex}
\begin{tabular}{cc}
  \scriptsize (a) $\sigma_1 = 1.2$ & \scriptsize (b) $\sigma_1 = 1.5$ \\[-1ex]	
  \includegraphics[width=.5\textwidth]{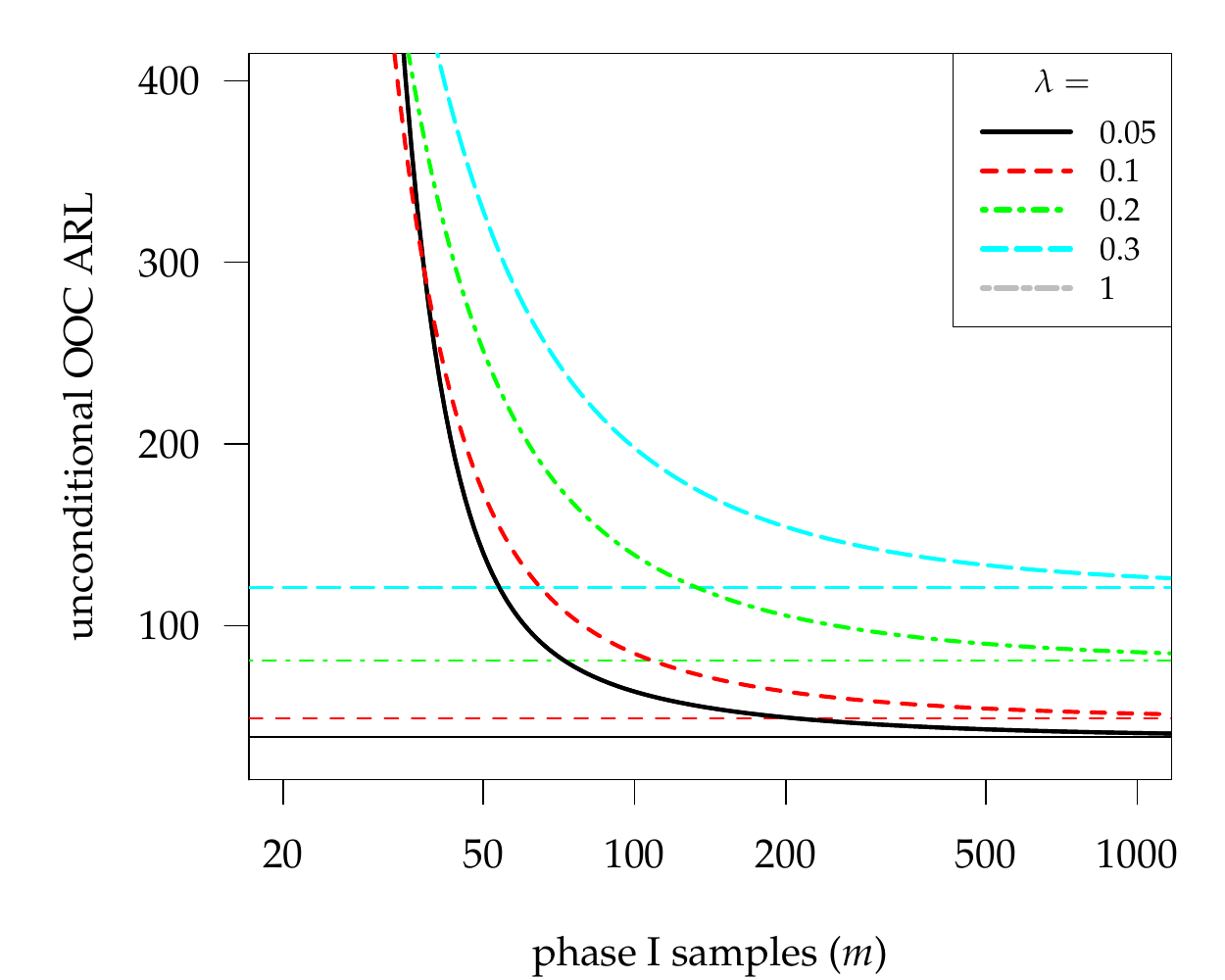} &
  \includegraphics[width=.5\textwidth]{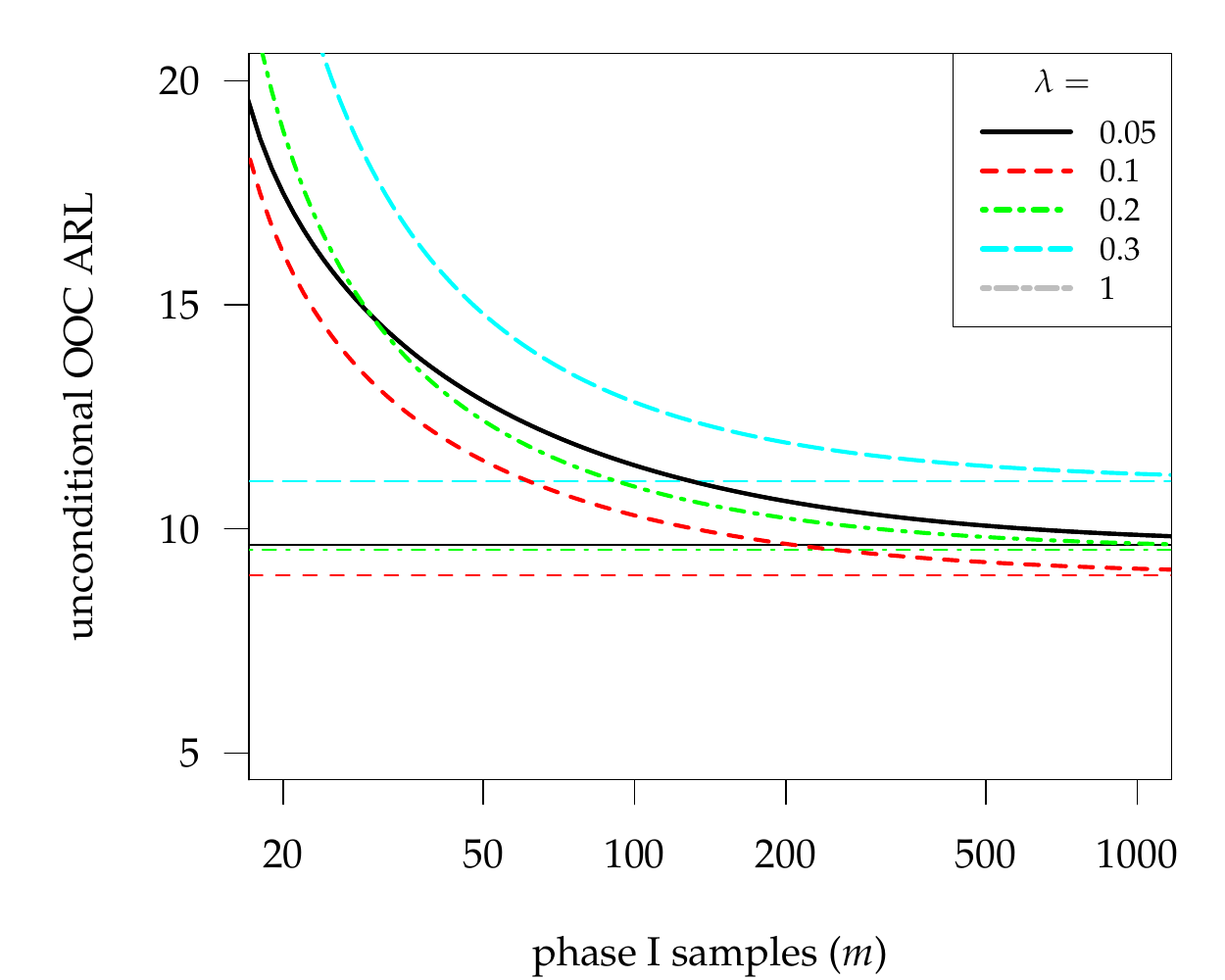} \\
  \scriptsize (c) $\sigma_1 = 0.8$ & \scriptsize (d) $\sigma_1 = 0.5$ \\[-1ex]
  \includegraphics[width=.5\textwidth]{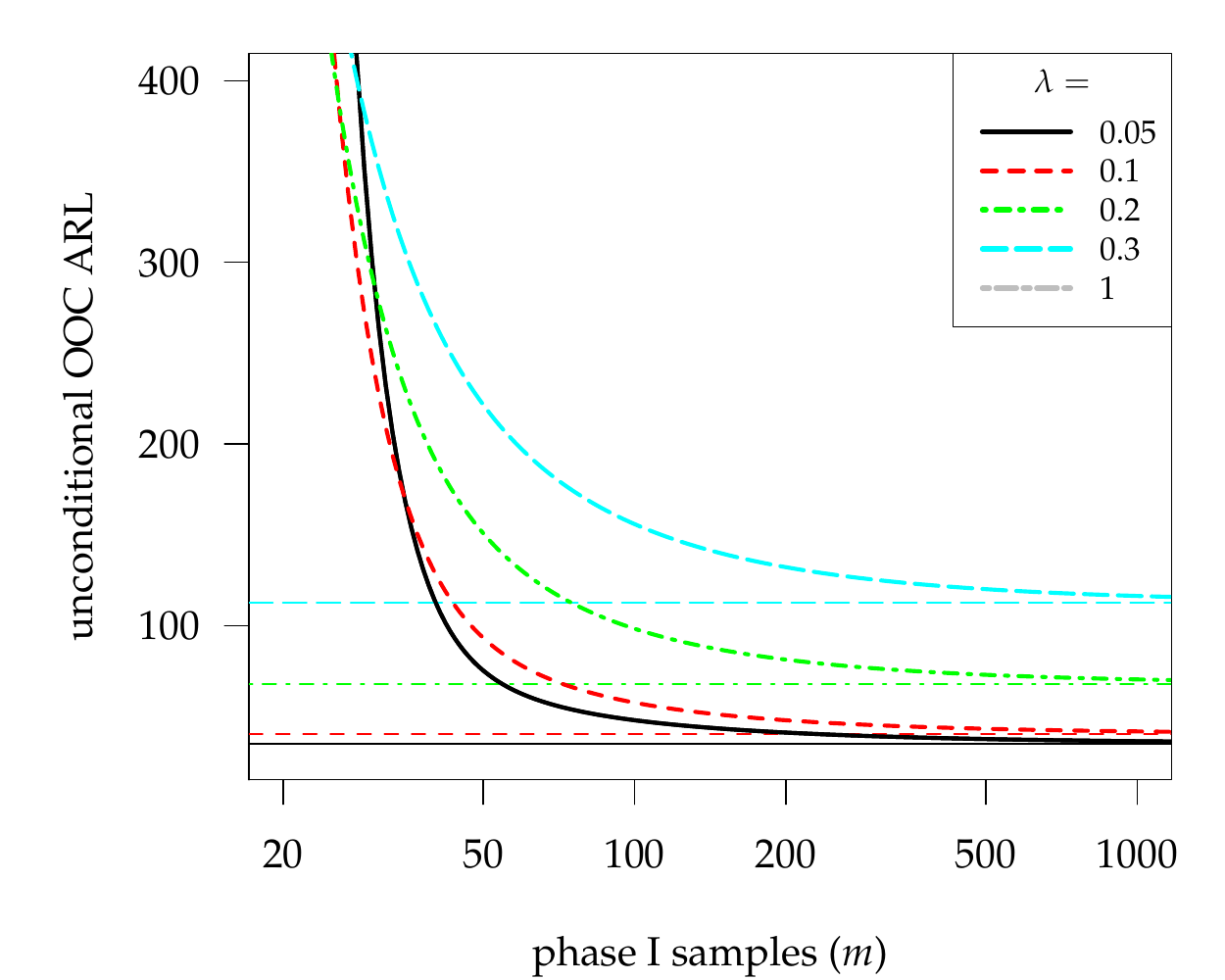} &
  \includegraphics[width=.5\textwidth]{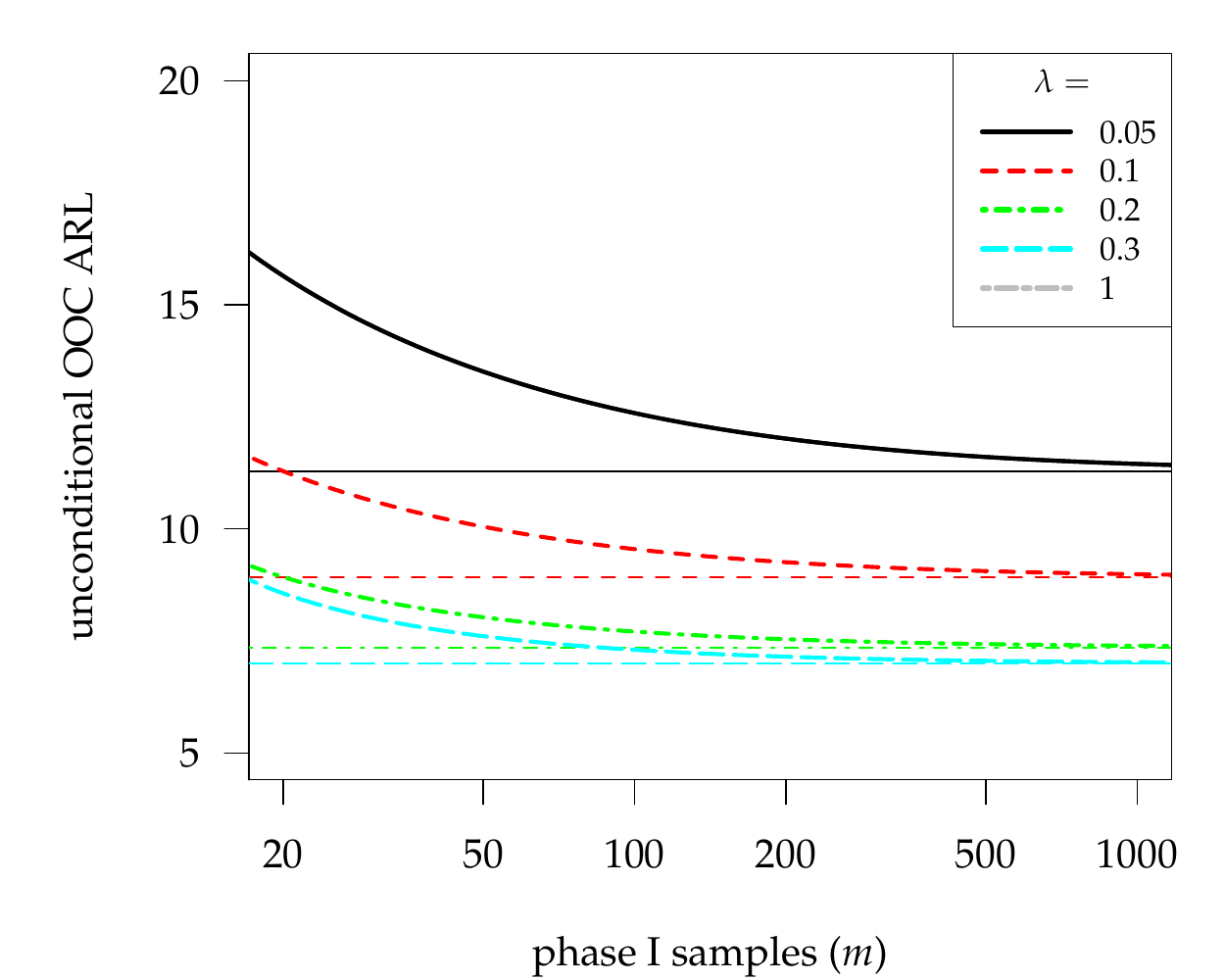} \\
\end{tabular} %
\caption{Unconditional OOC ARLs of the two-sided EWMA (various $\lambda$) $S^2$ ($n=5$) charts,
$P_\text{IC}(L\le 10^3) = 0.25$, phase I size $m=15, 16, \ldots, 1200$.}\label{fig:14}	
\end{figure}

Starting with the upper chart,
we present the unconditional OOC ARL values for
$\sigma_1 = 1.2$ and $\sigma_1 = 1.5$ in Figure~\ref{fig:13}
for the previously considered configurations.
In Figure~\ref{fig:13}(a), we detect two segments in the ARL profiles. For small values of $m < 50$,
we observe huge ARL values for all considered values of $\lambda$. In addition, the smaller the $\lambda$, the steeper
the curves, which completely changes the order of the analyzed control chart designs.
Given these patterns, it can be concluded that when attempting to detect small changes with an EWMA $S^2$ chart,
phase I samples with $m\ge 50$ are definitely needed. Namely, avoiding too many false alarms for small $m < 50$
leads inevitably to the delayed detection of small changes. Things look much better for the medium-sized change
$\sigma_1 = 1.5$ in Figure~\ref{fig:13}(b), where for all $\lambda$ and roughly all $m$,
the adjustment of the upper limit $c_u$ only mildly distorts the unconditional OOC ARL.
Moreover, the popular choices $\lambda = 0.1$ and $= 0.2$ produce overall decent ARL levels, indicating 
that a reasonable approach would be to recommend these two values, in general. Turning now to the
two-sided case in Figure~\ref{fig:14},
we confirm the judgments made for the upper schemes, where again small changes create problems for
$m <50$. The good news is that for the control chart user who is interested in detecting medium-sized and large
changes, the proposed adjustments of the control limits do not destroy the ability of the applied
EWMA charts to detect these changes. If flagging smaller changes is of concern, a larger phase I sample size is needed,
that is, $m \ge 50$, to obtain a detection performance that is comparable to that of the known parameter case.
It should be noted that the ARL values for the Shewhart chart are almost always too large to be displayed in
Figures~\ref{fig:13} and \ref{fig:14}.
The one and only exception in Figure~\ref{fig:13}(a) emphasizes that detecting a small variance increase 
in presence of an unknown IC variance level is difficult if only $m \le 40$ observations are available
to estimate the latter value.

\section{Conclusions}

In order to control the false alarm behavior of EWMA $S^2$ charts used for monitoring a normal variance,
we proposed an approach that widens the limits in a balanced way. The resulting control
chart design exhibits reasonable false alarm behavior while still being able to detect
medium-sized and large changes. To detect small changes, the phase I sample size must be increased
to $m \ge 50$ to achieve a performance that is comparable to the known parameter case.
Moreover, we believe that the notion that we are calibrating for a certain
false alarm probability $\alpha$ within a given number of control chart values (the chart horizon $\bar l$)
is easier to communicate to the statistical process monitoring community than declaring that one guarantees with
probability $1 - \alpha$
that the random conditional IC ARL is at least some nominal value, which corresponds more or less directly to $\bar l$
anyway. In addition, we recommend that $\lambda = 0.1$ or $= 0.2$ be used for setting up
a reasonable EWMA $S^2$ chart.
Finally, it should be noted that we have prepared an \texttt{R}
package (available from CRAN: \url{https://cran.r-project.org/})  that contains the functions needed to calculate the unconditional RL quantiles
and ARL values as well as the control charts limits (including their adjustments for a given phase I size $m$).
Some examples are given in the Appendix.
Moreover, the shiny app \url{https://kassandra.hsu-hh.de/apps/knoth/s2ewmaP/} provides a more convenient access.



\bibliographystyle{unsrtnat}
\bibliography{/home/knoth/common/references/sk}   

\appendix

\section{Software implementation}

The functions utilized throughout the paper are implemented in the \textsf{R} package \texttt{spc}.
For most of the figures and tables, the corresponding \textsf{R} code is provided as supplementary material
to this contribution. Here, some basic functions (\texttt{sewma.***.prerun()}) are described as follows.

\begin{verbatim}
install.packages("spc") # in case you have not installed the package so far

library(spc) # load the package

# configuration
df <- 5 - 1 # resulting degrees of freedom (for sample size 5)
m <- 50 # phase I size
lambda <- 0.2 # EWMA smoothing constant
ellbar <- 1000 # chart horizon
alpha <- 0.25 # false alarm probability

# control limits
clu <- sewma.q.crit.prerun(lambda, ellbar, alpha, df, m*df, sided="upper")
clu <- round(clu, digits=4)
print(clu)

# IC behavior
lMAX <- 15000 # right end of support of run length cdf
CDF <- 1 - sewma.sf.prerun(lMAX, lambda, 0, clu[2], 1, df, m*df, sided="upper")
plot(1:lMAX, CDF, type="l", xlab="l", ylab="P(L<=l)", log="x")
abline(v=ellbar, h=alpha, lty=2, col="grey")
points(ellbar, alpha, pch=19, col="grey")

# OOC behavior
SIGMA <- 1.5
CDF <- 1 - sewma.sf.prerun(lMAX, lambda, 0, clu[2], SIGMA, df, m*df, sided="upper")
plot(1:lMAX, CDF, type="l", xlab="l", ylab="P(L<=l)", log="x")
abline(v=ellbar, h=alpha, lty=2, col="grey")
points(ellbar, alpha, pch=19, col="grey")

# some unconditional ARL values
L10 <- sewma.arl.prerun(lambda, 0, clu[2], 1, df, m*df, sided="upper")
L15 <- sewma.arl.prerun(lambda, 0, clu[2], SIGMA, df, m*df, sided="upper")
print( data.frame(L10, L15), digits=3 )

## output should be two figures and ...
>     cl     cu 
> 0.0000 2.1538 
>       L10  L15
> arl 47128 9.79
\end{verbatim}

\end{document}